\documentclass[10pt]{article}
\usepackage{amsmath}
\usepackage{graphicx}
\usepackage{color}
\usepackage{orcidlink}
\usepackage{float}
\usepackage{hyperref}
\hypersetup{colorlinks=true, linkcolor=blue, citecolor=blue, urlcolor=blue}
\usepackage{caption}
\usepackage{subcaption}
\usepackage{setspace}
\usepackage{multirow}
\usepackage{multicol}
\usepackage{amssymb}
\RequirePackage[numbers,sort&compress]{natbib}
\begin{document}
\baselineskip=20pt

\begin{center}
\LARGE{ Shadow and Thermodynamics of deformed Schwarzschild-AdS black hole with a Cloud of Strings embedded in Perfect
Fluid Dark Matter }
\end{center}

\vspace{0.2cm}

\begin{center}

{\bf Faizuddin Ahmed}$^{a}$\textnormal{\orcidlink{0000-0003-2196-9622}}
\footnote{\bf faizuddinahmed15@gmail.com}\\
{\it $^{1}$Department of Physics, The Assam Royal Global University, Guwahati, 781035, Assam, India}

\vspace{0.3cm}

{\bf Carlos E. Romero-Figueroa}$^{b}$\textnormal{\orcidlink{0000-0001-5548-7766}}
\footnote{\bf carlosed.romero@correo.nucleares.unam.mx} and {\bf Hernando Quevedo}$^{b,c,d}$\textnormal{\orcidlink{0000-0003-4433-5550}}
\footnote{\bf quevedo@nucleares.unam.mx}\\

{\it $^{b}$Instituto de Ciencias Nucleares, Universidad Nacional Autónoma de México, AP 70543, Mexico City, Mexico}\\
{\it $^{c}$Dipartimento di Fisica and Icra, Università di Roma “La Sapienza”, Roma, Italy}\\
{\it $^{d}$Al-Farabi Kazakh National University, Al-Farabi av. 71, 050040 Almaty, Kazakhstan}

\end{center}

\vspace{0.2cm}
\begin{abstract}
We investigate the optical and thermodynamic properties of a deformed Schwarzschild Anti-de Sitter (AdS) black hole coupled to a cloud of strings and embedded in perfect fluid dark matter. We analyze the photon sphere and the corresponding black hole shadow, determining how these observables are affected by the string cloud, dark matter distribution, and geometric deformation. The thermodynamic behavior is studied through both the quasi-homogeneous fundamental equation within the classical physical approach and the geometric framework of geometrothermodynamics (GTD), showing full consistency between the two descriptions. In the extended phase space, we examine the critical structure and phase behavior of the system. Our analysis reveals that neither the geometric deformation nor the dark matter parameter generates new phase transitions; criticality emerges only in the vicinity of the Reissner--Nordström--AdS solution (RN--AdS), where the deformation parameter effectively plays the role of an electric charge squared. These results clarify the interplay between matter distributions, geometric deformations, and the phase structure of AdS black holes.
\end{abstract}

\tableofcontents

\section{Introduction}

General Relativity (GR) has withstood more than a century of experimental and observational scrutiny, ranging from the explanation of Mercury’s perihelion precession to the direct detection of gravitational waves by the LIGO-Virgo-KAGRA collaboration and the imaging of black hole shadows by the Event Horizon Telescope (EHT) collaboration~\cite{aa6,aa7,aa8,aa9,aa10,aa11,aa12,aa13,aa14}. Despite these remarkable successes, GR faces well-known theoretical and observational challenges. In particular, it does not provide a fundamental explanation for the dark sector of the Universe-namely dark matter and dark energy-and it remains a classical theory that is not yet reconciled with the principles of quantum mechanics. Motivated by these limitations, numerous modifications and extensions of GR have been proposed over the past decades. Meanwhile, rapid advances in observational astrophysics have opened unprecedented opportunities to test gravity in previously inaccessible regimes. Future missions and next-generation detectors, including Cosmic Explorer, the Einstein Telescope, LISA, TianQin, and Taiji, are expected to significantly improve the sensitivity of gravitational-wave observations, thereby enabling stringent tests of gravity in the strong-field regime. In this context, black holes provide an exceptionally promising laboratory for probing possible deviations from GR and exploring the fundamental nature of gravitation.

Black hole thermodynamics plays a central role in the quest for a consistent theory of quantum gravity. The thermodynamic interpretation of black holes was first proposed by Bekenstein in 1973~\cite{aa1}, who argued that if black holes possessed zero entropy, the total entropy of the Universe would decrease when matter carrying entropy falls into them, thereby violating the second law of thermodynamics. To avoid this inconsistency, he suggested that black holes must themselves carry entropy proportional to the area of their event horizon. Shortly thereafter, it was shown in~\cite{aa2} that black holes are not completely black but emit thermal radiation due to quantum effects near the horizon, a phenomenon now known as Hawking radiation. In addition, Hawking demonstrated that the surface area of a black hole event horizon cannot decrease in any classical process~\cite{aa3}, a result strikingly analogous to the second law of thermodynamics. These insights established a profound connection between gravity, thermodynamics, and quantum theory, culminating in the Bekenstein-Hawking area law, which relates the entropy $S$ of a black hole to the area $A$ of its event horizon via $S=\frac{A}{4}$, implying that black hole entropy scales with area rather than volume~\cite{aa3,aa4}. 

Further developments revealed an even richer thermodynamic structure. In particular, Kastor \textit{et al.}~\cite{aa15} proposed that, in asymptotically AdS spacetime, the cosmological constant should be interpreted as a thermodynamic variable in order to consistently generalize the Smarr relation. Within this extended phase space framework, the cosmological constant $\Lambda$ is identified with the thermodynamic pressure through $P=-\frac{\Lambda}{8\pi}$, and the black hole mass is reinterpreted as enthalpy rather than internal energy. This perspective gives rise to a remarkably rich phase structure: AdS black holes exhibit phase transitions between small and large black holes that closely resemble the liquid-gas transition of a Van der Waals fluid (VdW), with analogous equations of state, critical behavior, and critical exponents~\cite{aa16,aa17,aa18,aa19,aa20}. These results have motivated extensive studies of black hole chemistry and critical phenomena within extended gravitational thermodynamics.

Letelier~\cite{PSL1979} first introduced the concept of a cloud of strings in 1979, which may be regarded as a one-dimensional analogue of a cloud of dust. A black hole surrounded by a string cloud can be interpreted as a purely energetic field configuration, where the radial distribution of strings is balanced by an effective negative pressure that counteracts the inward gravitational pull~\cite{aa21}. One notable example in four dimensions is the global monopole configuration. The presence of a string cloud acts as an additional source of gravity, modifying the Schwarzschild solution and consequently altering the spacetime geometry~\cite{aa22}. In particular, it leads to an increase in the radius of the event horizon compared to the standard Schwarzschild case. An important outcome of studying Einstein’s field equations in the presence of a string cloud is that relativistic string distributions can provide useful phenomenological models for gravitational interactions. Studies of black hole solutions in the presence of a cloud of strings within the framework of Einstein gravity, as well as in various modified gravity theories, have been extensively investigated in the literature \cite{CS1,CS2,CS3,CS4,CS5,CS6,CS7,CS9,CS10,CS11}. 

Numerous astrophysical and cosmological observations provide compelling evidence for the existence of dark matter (DM), a non-luminous and non-baryonic component of the Universe. This component plays a fundamental role in the formation and evolution of large-scale structures and in shaping the cosmic web observed today. Current cosmological estimates indicate that only about $5\%$ of the total energy content of the Universe consists of ordinary (baryonic) matter, while the remainder is composed of dark matter and dark energy \cite{aa5,pp1,pp2}. Although dark matter neither emits, absorbs, nor reflects electromagnetic radiation, its presence is inferred indirectly through its gravitational effects \cite{pp3,pp4}. One of the earliest and most persuasive pieces of evidence arises from the rotation curves (RCs) of spiral galaxies \cite{pp5,pp6,pp7}, which remain approximately flat at large radial distances from the galactic center, in clear contradiction with predictions based solely on luminous matter. Moreover, gravitational lensing observations-both weak and strong-indicate mass distributions that substantially exceed those attributable to visible matter alone \cite{pp8,pp9}. Additional support for the dark matter paradigm comes from measurements of cosmic microwave background anisotropies \cite{pp10,pp11} and studies of large-scale structure formation \cite{pp12}. 

Despite the strong gravitational evidence supporting its existence, the fundamental nature and microscopic properties of dark matter remain unknown, rendering it one of the most significant open problems in contemporary physics. Among the various theoretical models proposed to describe dark matter, the perfect fluid dark matter (PFDM) model has attracted particular attention. In this framework, dark matter is modeled as a perfect fluid satisfying the equation of state $p/\rho = \epsilon$, which leads to a static spherically symmetric spacetime characterized by either a power-law or logarithmic metric function \cite{pp3,MHL2012,VVK}. The influence of PFDM on black hole spacetimes has been extensively investigated in the literature, including its effects on gravitational deflection angles, black hole shadows, and thermodynamic properties in various geometrical configurations \cite{kk1,kk2,kk3,kk4,kk6,kk7,kk8,kk9,kk10,kk11,kk12,kk13,kk14,kk15,kk16,kk17}.

In the current study, we investigate a static, spherically symmetric deformed Schwarzschild black hole with a cloud of strings (CS) and embedded in a perfect fluid dark matter background. We focus on the black hole's optical properties, as understanding how external matter distributions influence observables is of fundamental importance. Such effects can leave characteristic imprints on the photon sphere, black hole shadow, and related optical signatures. In particular, we analyze how the cloud of strings, PFDM, and deformation parameters modify these observables, providing insight into potential astrophysical and cosmological signatures. In addition to the optical analysis, we explore the thermodynamic behavior of this class of black holes using both the classical approach, based on the fundamental thermodynamic equation, and the geometric framework of geometrothermodynamics. Working within the extended phase space formalism, we also investigate the critical behavior and phase structure, highlighting how surrounding matter distributions affect the thermodynamic stability and phase transitions of AdS black holes. Overall, this study aims to provide a comprehensive understanding of the interplay between external matter, spacetime geometry, and black hole thermodynamics, with implications for both theoretical models and observational signatures.

\section{Spherically Symmetric deformed AdS BH with CS and PFDM}

In this section, we construct the background spacetime geometry of a static, spherically symmetric deformed Schwarzschild black hole \cite{MRK2023} endowed with a CS and embedded in a PFDM background. It is worth noting that various aspects of this deformed Schwarzschild-AdS black hole have been previously investigated in different contexts, including its thermodynamic properties \cite{DJG2025}, geodesic motion and scalar perturbations in the presence of a global monopole surrounded by a quintessence field \cite{FA2025}, holographic Einstein rings \cite{JYG2025}, observational signatures from anisotropic accretion disks \cite{DL2025}, and geodesic structure and thermodynamics in the presence of a cloud of strings \cite{PDU2025}. 

Building upon these studies, we examine how quantum deformation, PFDM, and the cloud of strings modify the spacetime curvature and consequently influence key astrophysical observables, such as the photon sphere, black hole shadow, and photon trajectories. In addition, we perform a comprehensive analysis of the thermodynamic properties of the system, exploring how the combined effects of quantum deformation, PFDM, and CS alter the thermodynamic behavior, stability, and phase structure of the black hole. 

The energy-momentum of perfect fluid dark matter is given by \cite{MHL2012}
\begin{equation}
T^{\text{DM}}_{\,\,\mu\nu} = \mathrm{diag}\left(\frac{\lambda}{8\pi r^3},\, \frac{\lambda}{8\pi r^3},\, -\frac{\lambda}{16\pi r^3},\, -\frac{\lambda}{16\pi r^3}\right).
\label{tensor}
\end{equation}
Here $\lambda$ is a constant that takes real values, called PFDM parameter.

To include string-like objects, we consider Nambu-Goto action given by \cite{PSL1979}
\begin{equation}
    S_{\rm CS}=\int \sqrt{-h}\,\mathcal{M}\,d\lambda^0\,d\lambda^1=\int \mathcal{M}\sqrt{-\frac{1}{2}\,\Sigma^{\mu \nu}\,\Sigma_{\mu\nu}}\,d\lambda^0\,d\lambda^1,\label{act1}
\end{equation}
where $\mathcal{M}$ is the dimensionless constant which characterizes the string, ($\lambda^0\,\lambda^1$) are the timelike and spacelike coordinate parameters, respectively \cite{JLS1960}. $h$ is the determinant of the induced metric of the strings world sheet given by $h_{ab}=g_{\mu\nu}\frac{ \partial x^\mu}{\partial \lambda^a}\frac{ \partial x^\nu}{\partial \lambda^b}$.  $\Sigma^{\mu\nu}=\epsilon^{ab}\frac{ \partial x^\mu}{\partial \lambda^a}\frac{ \partial x^\nu}{\partial \lambda^b}$ is bivector related to string world sheet, where $\epsilon^{ ab}$ is the second rank Levi-Civita tensor which takes the non-zero values as $\epsilon^{ 01} = -\epsilon^{ 10} = 1$.

The energy-momentum tensor of string-like objects is given by
\begin{equation}
   T_{\mu\nu}^{\rm CS}=2 \frac{\partial}{\partial g_{\mu \nu}}\mathcal{M}\sqrt{-\frac{1}{2}\Sigma^{\mu \nu}\,\Sigma_{\mu\nu}} =\frac{\rho^{\rm CS} \,\Sigma_{\alpha\nu}\, \,\Sigma_{\mu}^\alpha }{\sqrt{-h}}, \label{act2}
 \end{equation}
where $\rho^{\rm CS}$ is the proper density of the string cloud. The energy-momentum tensor components are given by
\begin{equation}
    T^{t\,(\rm CS)}_{t}=\rho^{\rm CS}=\frac{\gamma}{r^2}=T^{r\,(\rm CS)}_{r},\quad T^{\theta\,(\rm CS)}_{\theta}=T^{\phi\,(\rm CS)}_{\phi}= 0,\label{act3}
\end{equation}
Here $\gamma$ is a constant associated with strong-like objects.

Thereby, incorporating both cloud of strings and perfect fluid dark matter, the metric corresponding to deformed AdS black hole is described by the following line element
\begin{equation}
    ds^2 = -f(r)\,dt^2 + \dfrac{dr^2}{f(r)} + r^2 (d\theta ^2 + \sin ^2{\theta }\,d\varphi ^2),\label{metric}
\end{equation}
where the lapse function is given by
\begin{align}
     f(r) = 1-\gamma-\frac{2 M}{r}+\alpha \,\left[\frac{\beta ^2}{3r(\beta +r)^3}+\frac{1}{(\beta+r)^2}\right]-\frac{\Lambda}{3}\,r^2+\frac{\lambda}{r}\ln\!\frac{r}{|\lambda|},\label{function}
\end{align}
Here $M$ represent mass of the black hole, $\gamma$ is the string cloud parameter,  $\lambda$ is the PFDM parameter, $\Lambda$ is the cosmological constant and $\alpha, \beta$ are free positive constant parameters with $\beta$ being length dimension. The sign of the parameter $\gamma$ must satisfy $\gamma>0$ to ensure positive string tension density and a consistent geometric deficit. Values $\gamma>1$ may lead to pathological asymptotics. For the PFDM parameter, $\lambda<0$ is typically required to satisfy the energy conditions and guarantee a physically acceptable dark matter profile.

\begin{itemize}
    \item When $\alpha=0$, the metric function simplifies as
    \begin{align}
     f(r) = 1-\gamma-\frac{2 M}{r}-\frac{\Lambda}{3}\,r^2+\frac{\lambda}{r}\ln\!\frac{r}{|\lambda|}.\label{function1}
\end{align}
In that case, the metric Eq.~\eqref{metric} is the welll-known Letelier AdS black hole with perfect fluid dark matter \cite{AS2024}.

\item When $\beta=0$, the metric function simplifies as
\begin{align}
     f(r) = 1-\gamma-\frac{2 M}{r}+\frac{\alpha}{r^2}-\frac{\Lambda}{3}\,r^2+\frac{\lambda}{r}\ln\!\frac{r}{|\lambda|}.\label{function2}
\end{align}
Therefore, $\alpha=Q^2$ can be interpreted as the electric charge of black hole. In this particular case, the metric Eq. (\ref{metric}) is the Reissner-Nordstrom-AdS (RN-AdS) black hole with a cloud of strings and PFDM \cite{Singh2025}. Moreover, the solution reduces to pure RN-AdS BH for $\beta=0,\,\gamma=0=\lambda$, Letelier BH in the limit of $\alpha=0=\beta,\,\lambda=0$ and Schwarzschild-AdS BH for \(\alpha=0=\beta,\,\lambda=0=\gamma\). 

\begin{figure}[ht!]
 \centering
 \includegraphics[width=0.45\linewidth]{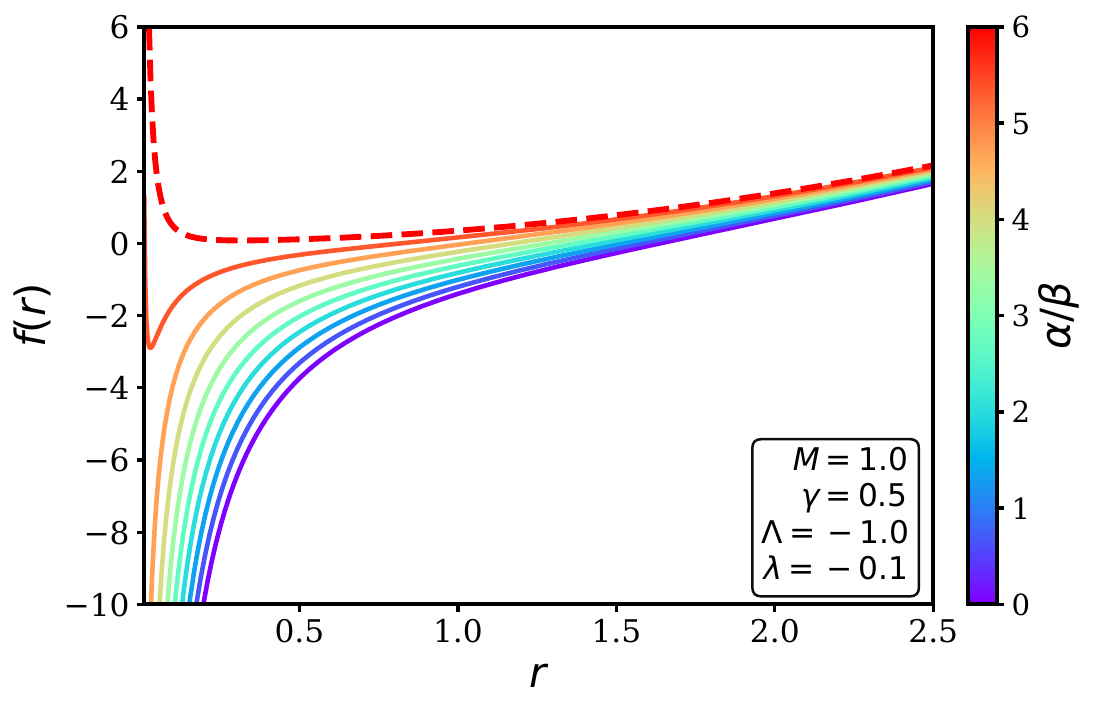}\qquad
 \includegraphics[width=0.48\linewidth]{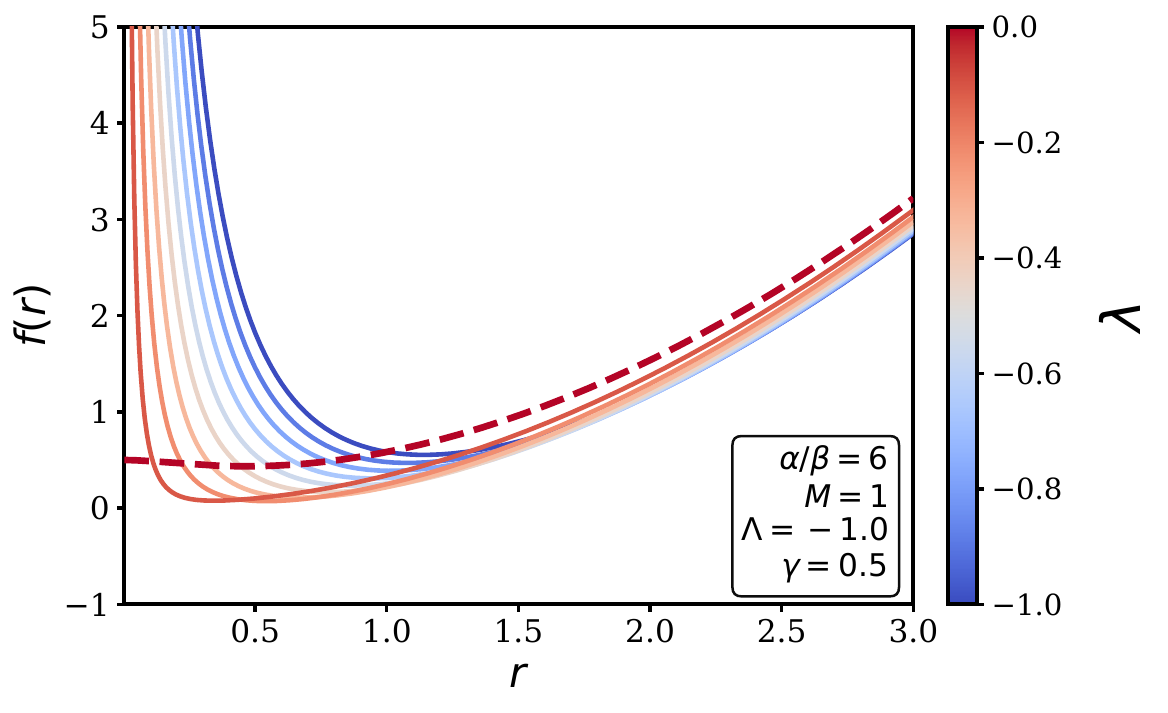}\\
 (i) \hspace{8cm} (ii)\\
 \includegraphics[width=0.48\linewidth]{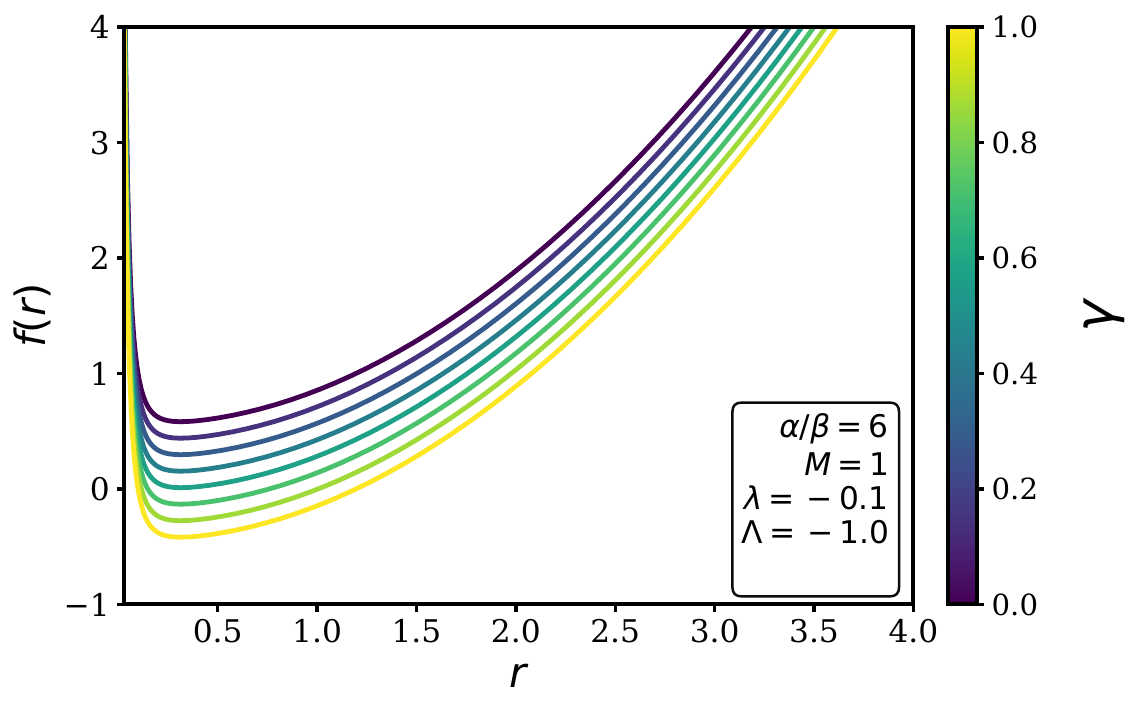}\\
 (iii)
\caption{Metric function $f(r)$. (i) Profile plot for varying $\alpha/\beta$. For sufficiently large values of the ratio $\alpha/\beta$,
naked singularity configurations arise (dashed curve). (ii) Profile plot for varying $\lambda$. We observe that a regular black hole is recovered only for $\alpha/\beta=6M$ and $\lambda \to 0$ (dashed curve). (iii) Profile plot varying $\gamma$, showing a global shift of the metric function.}
\label{metric function}
\end{figure}

\end{itemize}
\noindent
To analyze the behavior of the metric function near the origin, we consider the expansion for $r \to 0$. The metric function behaves as
\begin{align}
f(r)\Big|_{r\to 0} \sim \frac{\lambda}{r}\ln r - \frac{2M_{\rm eff}}{r} + 1-\gamma + \mathcal{O}(r^2).
\end{align}
The space-times near horizon acquires and effective mass 
\begin{align}
M_{\rm eff}=2M+\lambda \ln|\lambda|-\frac{\alpha}{3\beta}. \label{regular condition}
\end{align}

The coefficient of the $1/r$ term can be eliminated by imposing the fine-tuning condition $M_{\rm eff}=0$. However, even under this condition, the logarithmic contribution $\lambda/r\ln r$ remains divergent as $r \to 0$ for any $\lambda \neq 0$. Therefore, the presence of the PFDM parameter $\lambda$ introduces an unavoidable logarithmic divergence at the center as depicted in Fig. \ref{metric function}. This result indicates that the PFDM sector obstructs the existence of regular black hole solutions. In particular, from Eq.~\eqref{regular condition}, one concludes that regular configurations can only be recovered in the limit $\lambda \to 0$ together with the constraint $\alpha/\beta = 6M$, as illustrated in Fig.~\ref{metric function}(b). This behavior is fully consistent with the analysis reported in \cite{MRK2023}. Moreover, for large values of $\alpha/\beta$, one obtains $M_{\text{eff}}<0$, and the spacetime develops a naked singularity, as shown in Fig.~\ref{metric function}(a). Additionally, the parameter $\gamma$ produces a global shift of the metric function, affecting the horizon structure but leaving the central singularity unchanged Fig. \ref{metric function}(c).

\section{Black Hole Optics: Photon Sphere and Shadow}

Null geodesic motion is central to black hole physics because it governs how photons propagate through the curved spacetime surrounding a black hole. The properties of these null geodesics reveal key features such as the photon sphere, the black hole shadow, and the possible trajectories of light near the event horizon. As a result, they form the theoretical foundation for interpreting a wide range of astrophysical and observational phenomena associated with black holes \cite{SC1984,RMW1984}. The first horizon-scale images of the supermassive black holes \(M87^{*}\) \cite{aa9,aa10,aa11} and \(Sgr A^{*}\) \cite{aa12,aa13,aa14} obtained by the Event Horizon Telescope (EHT) has gained significant interest on black holes models in general relativity as well as modified gravity theories.

The space-time Eq.~(\ref{metric}) can be expressed as $ds^2=g_{\mu\nu} dx^{\mu} dx^{\nu}$, where the metric tensor $g_{\mu\nu}$ with $\mu,\nu=0,...,3$ is given by
\begin{equation}
    g_{\mu\nu}=\mbox{diag}\left(-f(r),\,\frac{1}{f(r)},\,r^2,\,r^2\, \sin^2 \theta\right),\label{bb1} 
\end{equation}

In this section, we investigate null geodesic motion using the Lagrangian formalism, which provides a powerful and systematic framework for analyzing photon trajectories in curved spacetime. Within this approach, the Lagrangian associated with a test particle moving in a gravitational background described by the metric tensor $g_{\mu\nu}$ is given by \cite{SC1984,RMW1984}
\begin{equation}
    \mathbb{L}=\frac{1}{2}g_{\mu\nu} \dot x^{\mu}\, \dot x^{\nu},\label{bb2}
\end{equation}
where dot represent ordinary derivative w. r. to $\lambda$, an affine parameter along geodesics.

Considering the geodesic motion on the equatorial plane defined by $\theta=\pi/2$, the Lagrangian density function Eq.~(\ref{bb2}) using Eq.~(\ref{bb1}) explicitly can be written as
\begin{equation}
    \mathbb{L}=\frac{1}{2}\left[-f(r) \dot t^2+\frac{1}{f(r)} \dot r^2+r^2\,\dot \phi^2\right].\label{bb3}
\end{equation}

One can see that the Lagrangian density function is independent of temporal coordinate $t$ and the angular coordinate $\theta$. Therefore, there exist two conserved quantities associated with these coordinates. These are called the energy of particles given by
\begin{equation}
    \mathrm{E}=f(r)\,\dot t,\label{bb4}
\end{equation}
And the conserved angular momentum by 
\begin{equation}
    \mathrm{L}=r^2\,\dot \phi,\label{bb5}
\end{equation}

\begin{figure}[ht!]
    \centering
    \includegraphics[width=0.45\linewidth]{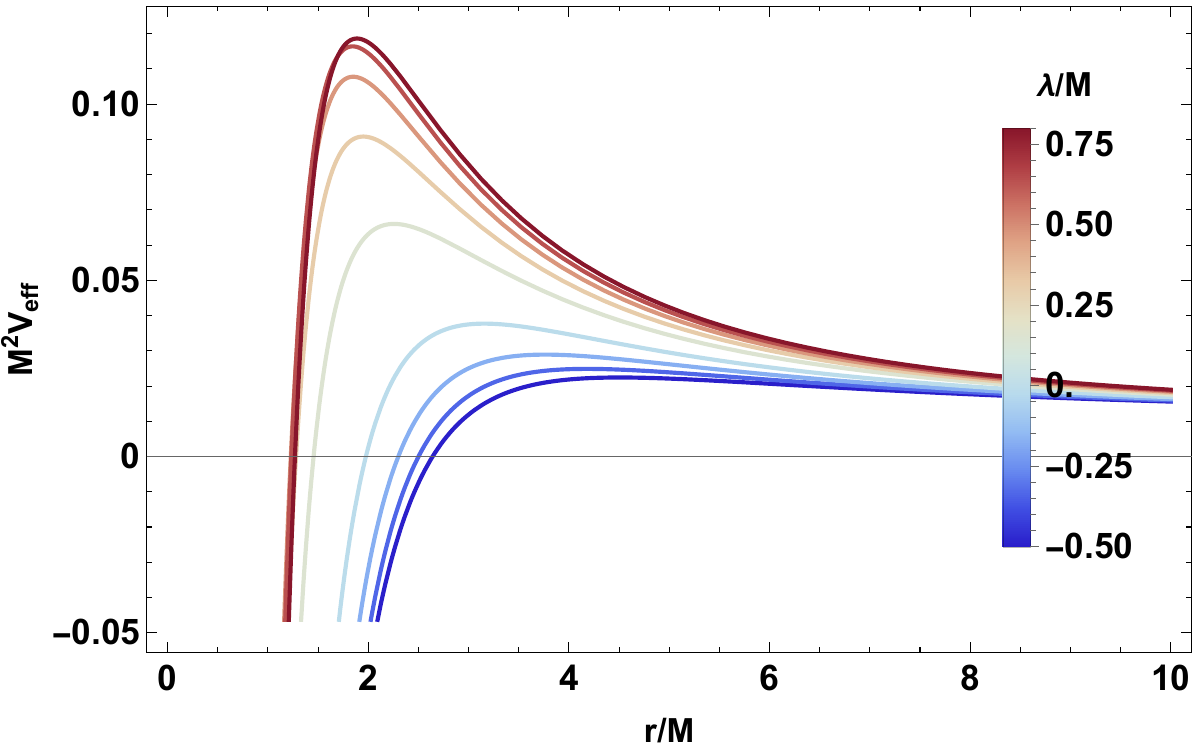}\qquad
    \includegraphics[width=0.48\linewidth]{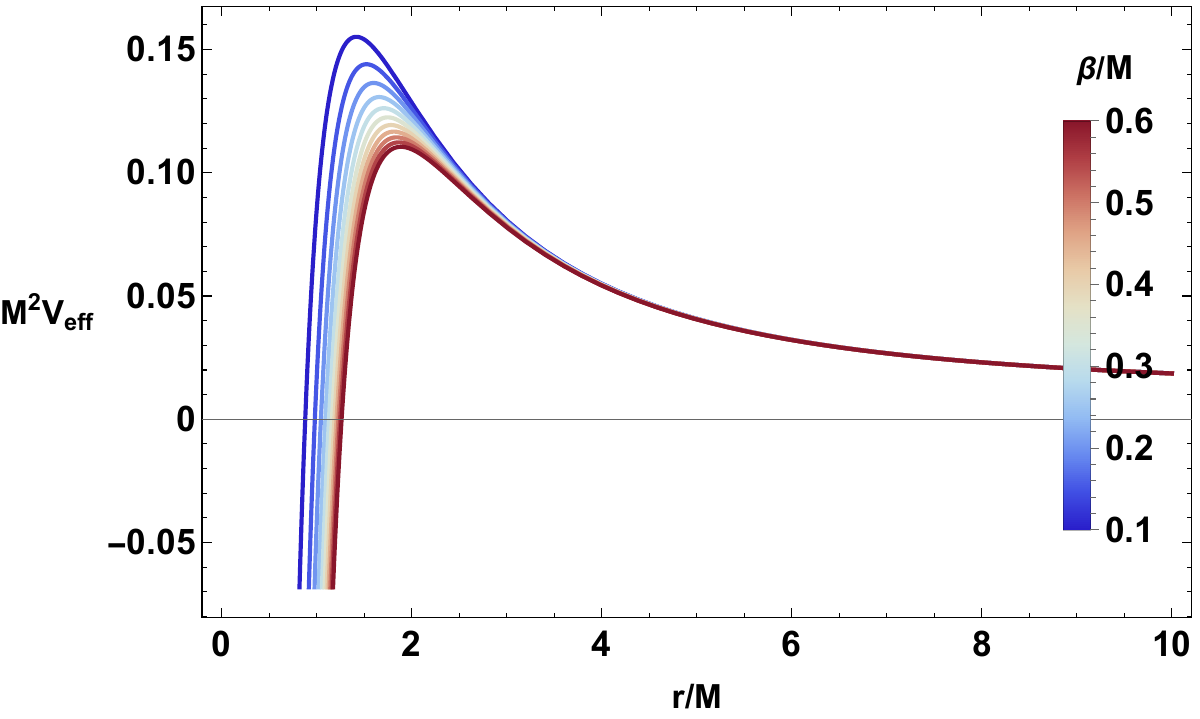}\\
    (i) $\gamma=0.1,\,\beta/M=0.5$ \hspace{6cm} (ii) $\gamma=0.1,\,\lambda/M=0.5$\\
    \includegraphics[width=0.48\linewidth]{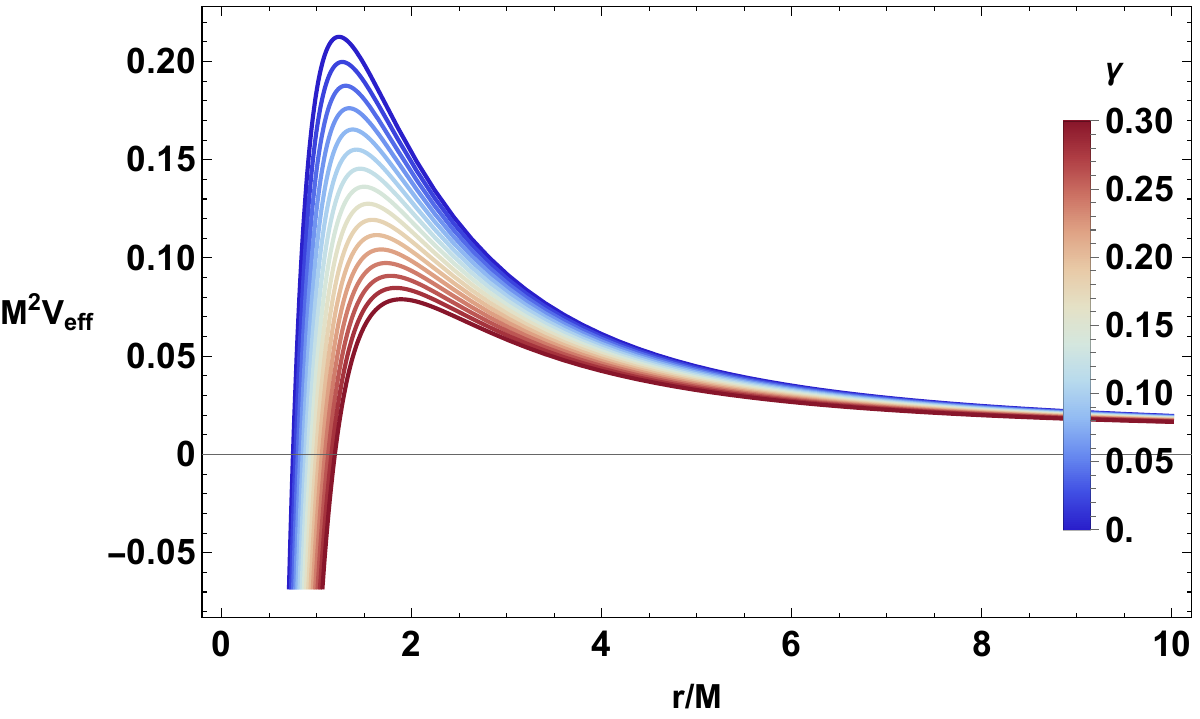}\\
    (iii) $\beta/M=0.1,\,\lambda/M=0.5$
    \caption{Behavior of the effective potential as a function of dimensonless radial distance $r/M$ for various values of $\lambda,\,\beta$ and $\gamma$. Here $\alpha/M=1,\,\mathrm{L}=1$.}
    \label{fig:1}
\end{figure}

Based on the above, the radial equation of motion for photon particles can be expressed as
\begin{equation}
    \dot r^2+V_{\rm eff}=\mathrm{E}^2,\label{bb6}
\end{equation}
which is equivalent to one-dimensional equation of motion of particles. Here $V_{\rm eff}$ is the effective potential of the system that governs the photon dynamics and is given by
\begin{equation}
    V_{\rm eff}=\frac{\mathrm{L}^2}{r^2}\,f(r)=\frac{\mathrm{L}^2}{r^2}\,\left[1-\gamma-\frac{2 M}{r}+\alpha\,\left\{\frac{\beta ^2}{3\,r\,(\beta +r)^3}+\frac{1}{(\beta+r)^2}\right\}-\frac{\Lambda}{3}\,r^2+\frac{\lambda}{r}\ln\!\frac{r}{|\lambda|}\right].\label{bb7}
\end{equation}
From the above expression, it is evident that the effective potential governing the photon dynamics is strongly influenced by the geometric and physical parameters of the spacetime. These parameters include the black hole mass $M$, the string cloud parameter $\gamma$, the perfect fluid dark matter parameter $\lambda$, the deformation parameters $(\alpha, \beta)$, as well as the cosmological constant $\Lambda$. Consequently, the behavior of null geodesics and the structure of photon orbits are highly sensitive to variations in these quantities.

In Fig.~\ref{fig:1}, we present the variation of the dimensionless quantity $M^{2}V_{\rm eff}$ as a function of the scaled radial coordinate $r/M$. The plots are obtained by varying the PFDM parameter $\lambda$, the deformation parameter $\beta$, and the cloud of strings parameter $\gamma$, while keeping the angular momentum $\mathrm{L}$ and the deformation parameter $\alpha$ fixed at unity. Panel~\ref{fig:1}(i) shows that an increase in the PFDM parameter $\lambda$ leads to a higher peak of the effective potential, indicating a strengthening of the potential barrier. Conversely, panels~\ref{fig:1}(ii)--(iii) demonstrate that increasing either the deformation parameter $\beta$ or the string cloud parameter $\gamma$ results in a reduction of the peak height. This behavior suggests that while PFDM enhances the gravitational potential barrier, the deformation and string cloud effects tend to weaken it.

Next, we turn to the study of circular null orbits, with particular emphasis on the photon sphere, and examine how the geometric parameters of the spacetime influence its properties. The photon sphere is a spherical hypersurface surrounding a black hole on which photons can move along unstable circular null geodesics. Its radius is highly sensitive to the underlying spacetime geometry and, consequently, to the black hole parameters.

For circular null geodesics, the conditions \(\dot{r} = 0,\, \ddot{r} = 0\) must be simultaneously satisfied. Using Eq.~(\ref{bb6}), these conditions lead to the following relations:
\begin{equation}
    \mathrm{E}^2=V_{\rm eff}=\frac{\mathrm{L}^2}{r^2}\,f(r).\label{bb8}
\end{equation}
And
\begin{equation}
    \frac{d}{dr}\left(\frac{f(r)}{r^2}\right)=0.\label{bb9}
\end{equation}

The relation Eq.~(\ref{bb9}) gives us the photon sphere radius $r=r_s$ satisfying the following polynomial equation as,
\begin{align}
1 - \gamma
- \frac{3M}{r}
+ \frac{3\lambda}{2r}\ln\!\left(\frac{r}{|\lambda|}\right)
- \frac{\lambda}{2r}
+ \alpha \left[
\frac{\beta^2(6r^2+7\beta r+2\beta^2)}
{6r(\beta+r)^4}
+
\frac{2\beta+3r}{2(\beta+r)^3}
\right]=0.\label{bb10}
\end{align}
The exact analytical solution of the above polynomial will give the photon sphere radius $r_s$. Noted that the exact analytical solution is quite a challenging due to higher order polynomial. However, one can determine the numerical results by selecting suitable values parameters.

\begin{table}[ht!]
\centering
\begin{tabular}{|c|c|c|c|c|c|}
\hline
$\beta/M (\downarrow) \backslash \gamma (\rightarrow )$ & 0.00 & 0.05 & 0.10 & 0.15 & 0.20 \\
\hline
0.0 & 2.94895 & 3.14841 & 3.36934 & 3.61567 & 0.54148 \\
0.1 & 2.99496 & 3.19052 & 3.40781 & 3.65075 & 3.92424 \\
0.2 & 3.03167 & 3.22459 & 3.43936 & 3.67988 & 3.95104 \\
0.3 & 3.06153 & 3.25261 & 3.46557 & 3.70431 & 3.97374 \\
0.4 & 3.08616 & 3.27592 & 3.48755 & 3.72497 & 3.99308 \\
\hline
\end{tabular}
\caption{Numerical solutions for $r_s/M$ for various $\beta$ and $\gamma$ values. Here $\alpha/M^2=1,\,\lambda/M=-0.10$.}
\label{tab:1}
\end{table}

\begin{table}[ht!]
\centering
\begin{tabular}{|c|c|c|c|c|c|}
\hline
$\beta/M (\downarrow) \backslash \gamma (\rightarrow )$ & 0.00 & 0.05 & 0.10 & 0.15 & 0.20 \\
\hline
0.0 & 3.12916 & 3.34146 & 3.57696 & 3.83994 & 4.13573 \\
0.1 & 3.17047 & 3.37926 & 3.61149 & 3.87141 & 4.16433 \\
0.2 & 3.20389 & 3.41024 & 3.64013 & 3.89781 & 4.18859 \\
0.3 & 3.23137 & 3.43597 & 3.66415 & 3.92015 & 4.20929 \\
0.4 & 3.25421 & 3.45753 & 3.68443 & 3.93916 & 4.22703 \\
\hline
\end{tabular}
\caption{Numerical solutions for $r_s$ for various $\beta$ and $\gamma$ values. Here $\alpha/M^2=1,\,\lambda/M=-0.15$.}
\label{tab:2}
\end{table}

\begin{figure}[ht!]
    \centering
    \includegraphics[width=0.45\linewidth]{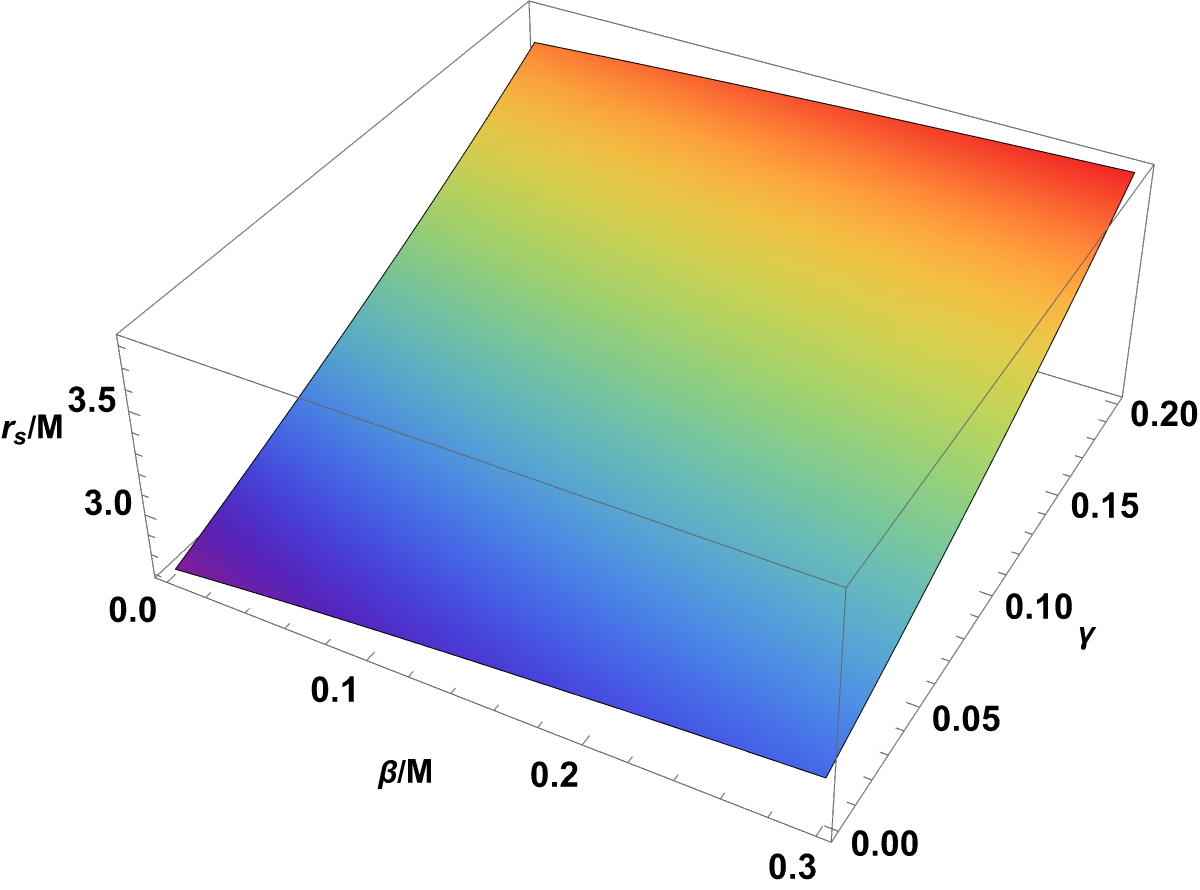}\qquad
    \includegraphics[width=0.45\linewidth]{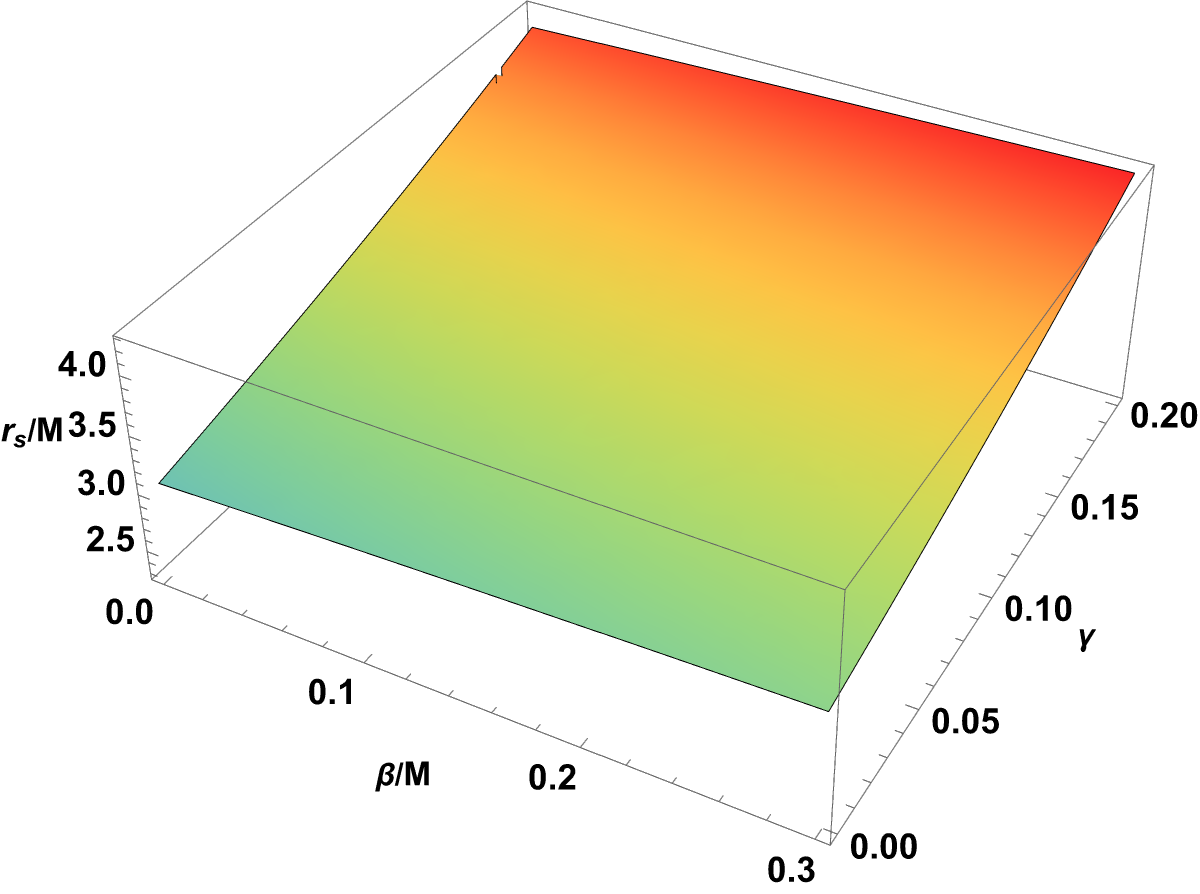}\\
    (i) $\lambda/M=-0.1$ \hspace{6cm} (ii) $\lambda/M=-0.2$\\
    \caption{Photon sphere radius as a function of $\beta/M$ and $\gamma$ for two values of $\lambda/M$. Here $\alpha=1$.}
    \label{fig:2}
\end{figure}

Tables~\ref{tab:1} and \ref{tab:2} present the numerical values of the photon sphere radius obtained by simultaneously varying the deformation parameter $\beta$ and the string cloud parameter $\gamma$ for two fixed values of the PFDM parameter, $\lambda/M = -0.10$ and $\lambda/M = -0.15$. From both tables, it is evident that for a given value of $\beta$, the photon sphere radius increases with increasing $\gamma$. A similar trend is observed for fixed $\gamma$, where an increase in the magnitude of the PFDM parameter $\lambda$ leads to a corresponding enlargement of the photon sphere radius. These results indicate that both the string cloud and PFDM parameters tend to expand the photon sphere.

Furthermore, Fig.~\ref{fig:2} illustrates the behavior of the scaled photon sphere radius $r_s/M$ as a function of the parameters $(\beta, \gamma)$ for two representative values, $\lambda/M = -0.10$ and $\lambda/M = -0.20$. The figure confirms the trends observed in the tables, clearly demonstrating how the combined effects of deformation and external matter distributions modify the size of the photon sphere.

Finally, we focus into the BH shadow showing the effects of geometric parameters. The black hole shadow corresponds to the collection of photon trajectories that are captured by the black hole and therefore do not reach a distant observer. It arises as a purely gravitational effect, governed entirely by the behavior of null geodesics in the strong–field region of the spacetime. It represents the apparent dark region on the observer’s sky formed by photons that asymptotically approach unstable circular orbits and eventually fall into the black hole.

The boundary of the shadow is determined by the critical impact parameter $b_c$ associated with the unstable photon sphere. This critical value separates two classes of null trajectories: photons with impact parameters smaller than the critical value ($b<b_c$) spiral inward and are captured by the black hole, while those with larger impact parameters ($b>b_c$) are scattered and escape to infinity. Consequently, the size and shape of the shadow encode detailed information about the spacetime geometry and depend sensitively on the black hole parameters and any additional fields or deformations present in the gravitational background.

The critical impact parameter at radius $r=r_s$ using Eq. (\ref{bb8}) is given by
\begin{equation}
    b_c=\frac{\mathrm{E}(r_s)}{\mathrm{L}(r_s)}=\frac{r_s}{\sqrt{f(r_s)}}=\frac{r_s}{\sqrt{1-\gamma-\frac{2 M}{r_s}+\alpha \,\left[\frac{\beta^2}{3\,r_s\,(\beta +r_s)^3}+\frac{1}{(\beta+r_s)^2}\right]+\frac{\lambda}{r_s}\ln\!\frac{r_s}{|\lambda|}}}.\label{critical}
\end{equation}

To determine the shadow radius, we examine the behavior of the metric function $f(r)$ at large distances. Noted that at large radial distance, the metric function without cosmological constant simplifies as
\begin{equation}
    \lim_{r \to \infty} f(r) = 1 - \gamma \neq 1,\label{bb11}
\end{equation}
This indicates that the function is asymptotically bounded rather than flat space. Therefore, the angular size of the radius for a static observer located at position $r_{O}$ is defined by \cite{Volker2022}
\begin{equation}
    \sin \vartheta_{\rm sh}=\frac{h(r_{s})}{h(r_O)},\qquad h^2(r)=\frac{r^2}{f(r)}.\label{bb12}
\end{equation}
Substituting the metric components, we find
\begin{equation}
   \sin \vartheta_{\rm sh}=\frac{r_s}{r_O}\,\sqrt{\frac{1-\gamma-\frac{2 M}{r_O}+\alpha \,\left[\frac{\beta ^2}{3\,r_O\,(\beta +r_O)^3}+\frac{1}{(\beta+r_O)^2}\right]-\frac{\Lambda}{3}\,r^2_O+\frac{\lambda}{r_O}\ln\!\frac{r_O}{|\lambda|}}{1-\gamma-\frac{2 M}{r_s}+\alpha \,\left[\frac{\beta^2}{3\,r_s\,(\beta +r_s)^3}+\frac{1}{(\beta+r_s)^2}\right]-\frac{\Lambda}{3}\,r^2_s+\frac{\lambda}{r_s}\ln\!\frac{r_s}{|\lambda|}}}.\label{bb13}
\end{equation}
Therefore, the shadow radius is given by (taking zero cosmological constant)
\begin{equation}
   R_{\rm sh} \simeq r_O\,\vartheta_{\rm sh}=r_s\,\sqrt{\frac{1-\gamma-\frac{2 M}{r_O}+\alpha \,\left[\frac{\beta ^2}{3\,r_O\,(\beta +r_O)^3}+\frac{1}{(\beta+r_O)^2}\right]+\frac{\lambda}{r_O}\ln\!\frac{r_O}{|\lambda|}}{1-\gamma-\frac{2 M}{r_s}+\alpha \,\left[\frac{\beta^2}{3\,r_s\,(\beta +r_s)^3}+\frac{1}{(\beta+r_s)^2}\right]+\frac{\lambda}{r_s}\ln\!\frac{r_s}{|\lambda|}}}.\label{bb14}
\end{equation}

For a distant observer, the shadow radius simplifies as,
\begin{equation}
   R_{\rm sh}=(1-\gamma)^{1/2}\,\frac{r_s}{\sqrt{1-\gamma-\frac{2 M}{r_s}+\alpha \,\left[\frac{\beta^2}{3\,r_s\,(\beta +r_s)^3}+\frac{1}{(\beta+r_s)^2}\right]+\frac{\lambda}{r_s}\ln\!\frac{r_s}{|\lambda|}}}=(1-\gamma)^{1/2}\,b_c.\label{bb15}
\end{equation}
From the above analysis, we observe that the shadow radius is differ from the critical impact parameter. This is because of the presence of string-like objects that produces an solid angle deficit.

\begin{table}[ht!]
\centering
\begin{tabular}{|c|c|c|c|c|c|}
\hline
$\beta/M (\downarrow) \backslash \gamma (\rightarrow )$ & 0.00 & 0.05 & 0.10 & 0.15 & 0.20 \\
\hline
0.0 & 5.1966 & 5.54672 & 5.93454 & 6.36696 & 1.06939 \\
0.1 & 5.26862 & 5.6125 & 5.99453 & 6.42158 & 6.90225 \\
0.2 & 5.32903 & 5.66852 & 6.04635 & 6.46938 & 6.94619 \\
0.3 & 5.38082 & 5.7171 & 6.09178 & 6.51172 & 6.98551 \\
0.4 & 5.42594 & 5.75981 & 6.13207 & 6.5496 & 7.02097 \\
\hline
\end{tabular}
\caption{Numerical values of the shadow radius $R_{\rm sh}/M$ for various $\beta$ and $\gamma$ values. Here $\alpha/M^2=1,\,\lambda=-0.10,\,r_O/M=100$}
\label{tab:3}
\end{table}

\begin{table}[ht!]
\centering
\begin{tabular}{|c|c|c|c|c|c|}
\hline
$\beta/M (\downarrow) \backslash \gamma (\rightarrow )$ & 0.00 & 0.05 & 0.10 & 0.15 & 0.20 \\
\hline
0.0 & 5.55464 & 5.92937 & 6.34507 & 6.80928 & 7.33141 \\
0.1 & 5.61912 & 5.98827 & 6.39879 & 6.85817 & 7.37579 \\
0.2 & 5.67403 & 6.03913 & 6.44577 & 6.90143 & 7.41550 \\
0.3 & 5.72164 & 6.08369 & 6.48735 & 6.94010 & 7.45132 \\
0.4 & 5.76350 & 6.12320 & 6.52453 & 6.97495 & 7.48387 \\
\hline
\end{tabular}
\caption{Numerical values of the shadow radius $R_{\rm sh}$ for various $\beta$ and $\gamma$ values. Here $\alpha/M^2=1,\,\lambda=-0.15,\,r_O/M=100$}
\label{tab:4}
\end{table}

\begin{figure}[ht!]
    \centering
    \includegraphics[width=0.45\linewidth]{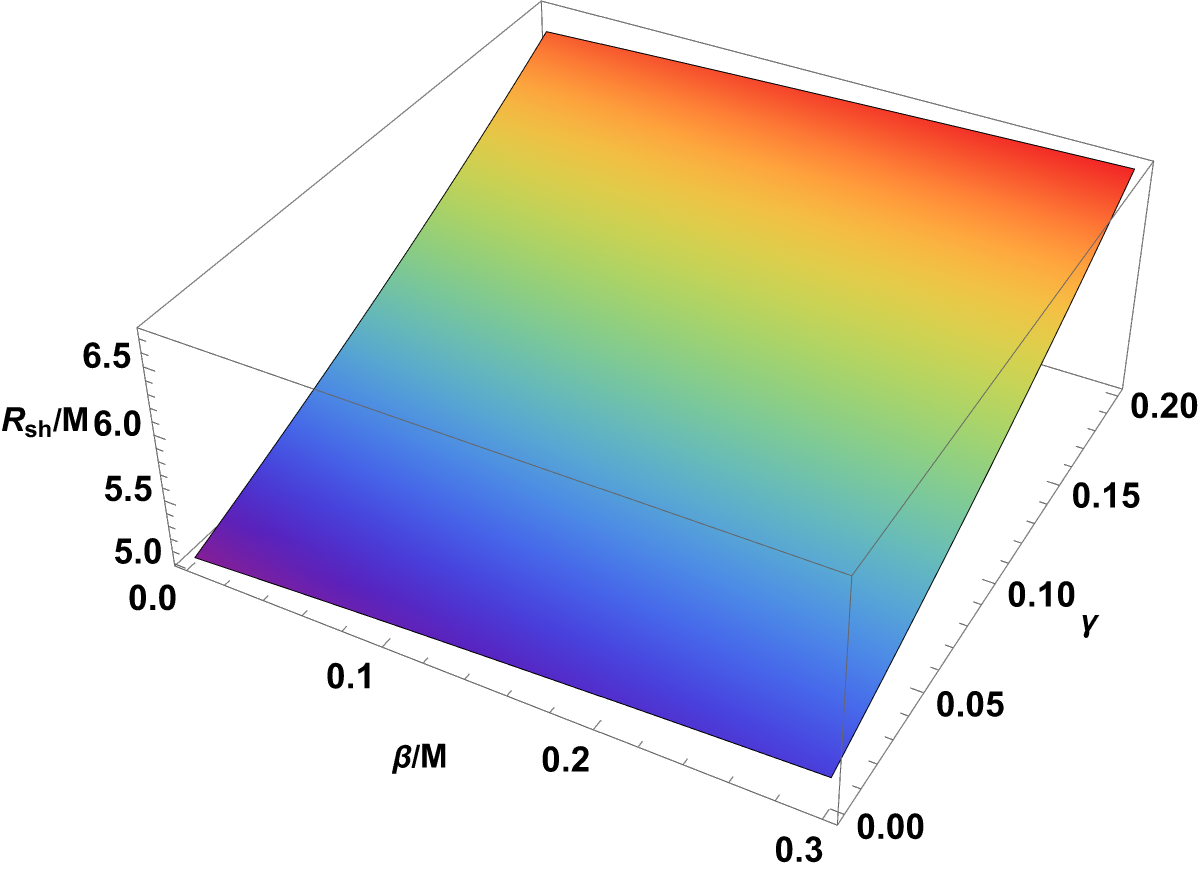}\qquad
    \includegraphics[width=0.45\linewidth]{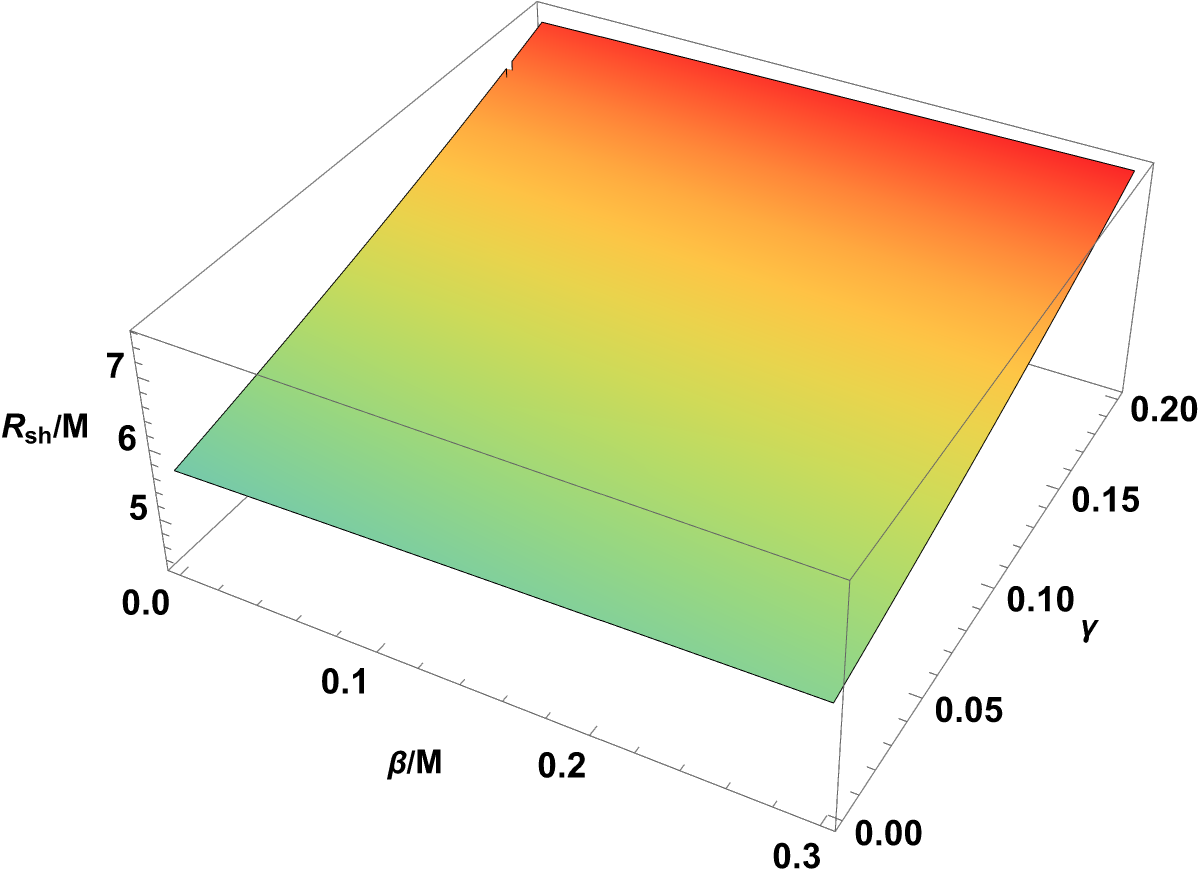}\\
    (i) $\lambda/M=-0.1$ \hspace{6cm} (ii) $\lambda/M=-0.2$\\
    \caption{Shadow radius as a function of $\beta/M$ and $\gamma$ for two values of $\lambda/M$. Here $\alpha/M=1,\,r_O/M=50$.}
    \label{fig:3}
\end{figure}

Tables~\ref{tab:3} and \ref{tab:4} present the numerical values of the shadow radius $R_{\rm sh}$ obtained by simultaneously varying the deformation parameter $\beta$ and the string cloud parameter $\gamma$ for two fixed values of the PFDM parameter, $\lambda/M = -0.10$ and $\lambda/M = -0.15$ with observer location $r_O/M=100$. From both tables, it is evident that for a given value of $\beta$, the shadow radius increases with increasing $\gamma$. A similar trend is observed for fixed $\gamma$, where an increase in the magnitude of the PFDM parameter $\lambda$ leads to a corresponding enlargement of the photon sphere radius. These results indicate that both the string cloud and PFDM parameters tend to expand the photon sphere.

Furthermore, Fig.~\ref{fig:3} illustrates the behavior of the scaled shadow radius $R_{\rm sh}/M$ as a function of the parameters $(\beta, \gamma)$ for two representative values, $\lambda/M = -0.10$ and $\lambda/M = -0.20$ with observer location $r_O/M=50$. The figure confirms the trends observed in the tables, clearly demonstrating how the combined effects of deformation and external matter distributions modify the size of the photon sphere.

\section{Thermodynamic Properties }

We begin this section by exploring the thermodynamics of the deformed AdS--black hole surrounded by a CS and PFDM. To achieve this, we first analyze the entropy of the system
\begin{equation}
S=\frac{A}{4}=\frac{1}{4}\,\lim_{r \to r_h}\,\int^{\pi}_{\theta=0} \int^{2\pi}_{\phi=0} \sqrt{g_{\theta\theta}\,g_{\phi\phi}}\, d\theta\,d\phi.\label{ff5}
\end{equation}
In our case at hand, we find
\begin{equation}
    S=\pi r^2_h.\label{entropy}
\end{equation}
The thermodynamic properties of the system are fully encoded in the
fundamental equation $M=M(S, P,\gamma,\lambda,\alpha,\beta)$, where $P=-\Lambda/8 \pi$ as usual in the context of extended black hole thermodynamics \cite{kubizvnak2017black}. 
This relation is obtained by imposing the horizon condition $f(r_h)=0$ and expressing the result in terms of the Bekenstein entropy Eq. \eqref{entropy}, and reads
\begin{equation}
M=
\frac{1}{6}\left[
\frac{3\sqrt{S}}{\sqrt{\pi}}(1-\gamma)
+
\frac{
\alpha\!\left(
\frac{3S}{\pi}
+
\frac{3\beta\sqrt{S}}{\sqrt{\pi}}
+
\beta^{2}
\right)
}{
\left(
\frac{\sqrt{S}}{\sqrt{\pi}}+\beta
\right)^{3}
}
+
\frac{8PS^{3/2}}{\sqrt{\pi}}
+
3\lambda
\ln\!\left(
\frac{\sqrt{S}}{\sqrt{\pi}\,|\lambda|}
\right)
\right]. \label{funda}
\end{equation}
The fundamental equation $M(X^i)$ defines a quasi-homogeneous
thermodynamic system of degree $w_M=1/2$.
This property follows from the dimensional scaling behavior of the
entropy and coupling constants
$X^i=\{S,\Lambda,\gamma,\lambda,\alpha,\beta\}$, which transform as
$X^i\to\mu^{w_i}X^i$.
Under this generalized scaling, the parameters obey
\begin{equation}
S\to \mu\, S,\qquad
\Lambda\to \mu^{-1}\Lambda,\qquad
\gamma\to \mu^{0}\gamma,\qquad
\lambda\to \mu^{1/2}\lambda,\qquad
\alpha\to \mu\,\alpha,\qquad
\beta\to \mu^{1/2}\beta, \label{wi}
\end{equation}
where $\mu$ is an arbitrary scaling parameter. As a consequence, the mass transforms according to
\begin{equation}
M(\mu^{w_i}X^i)=\mu^{1/2}M(X^i).
\end{equation}
Moreover, using the conjugate thermodynamic parameters $\mathcal{Y}_i\equiv\partial M/\partial X^i$, Euler’s theorem then yields the generalized Smarr relation \cite{romero2024extended1}
\begin{equation}
M=2\sum_i w_i X^i\mathcal{Y}_{i}=2TS-2PV+2\Pi_\alpha \alpha+\Pi_\beta \beta+\Pi_\lambda \lambda ,\label{smarr}
\end{equation}
where the conjugate thermodynamic variables are
\begin{equation}
T\equiv\left(\frac{\partial M}{\partial S}\right)_{P,\Tilde{X}^{j}},
\qquad
V\equiv\left(\frac{\partial M}{\partial P}\right)_{S,\Tilde{X}^{j}},\qquad 
\Pi_{j}\equiv
\left(\frac{\partial M}{\partial X^{j}}\right)_{S,P,\Tilde{X}^{k\neq j}},
\qquad
\Tilde{X}^{j}=\{\lambda,\alpha,\beta\}. \label{EoS}
\end{equation}
Promoting the coupling constants of the theory to thermodynamic
parameters \cite{romero2024extended1}, the extended first law\footnote{%
Since the  parameter $\gamma$ is dimensionless, it carries 
zero scaling weight in the quasi-homogeneous framework. Therefore, it does 
not appear in the Smarr relation and no $d\gamma$ term arises in the first 
law derived from it.} of black hole thermodynamics follows directly from Eq. \eqref{smarr} and can be written as
\begin{equation}
dM=\sum_i \mathcal{Y}_i\,dX^i
= T\,dS+V\,dP+\sum_j\Pi_{j}d\Tilde{X}^j, \label{first law}
\end{equation}
Additionally, since the fundamental potential is quasi-homogeneous, it leads to a generalized Gibbs--Duhem relation of
the form \cite{romero2026quasi}
\begin{equation}
\sum_{i=1}^{n} 
w_i 
\left[
\left(1 - \frac{w_M}{w_i} \right)
dM
+ 
X^i \, d\mathcal{Y}_i
\right]
= 0 .
\end{equation}
In the particular case in which the fundamental relation $M$ is homogeneous of degree $w_M=1$, one recovers the standard Gibbs--Duhem relation,
\begin{equation}
\sum_i w_i X^i \, d\mathcal{Y}_i = 0 \ .
\end{equation}
For the system under consideration, however, the black hole is quasi-homogeneous of degree $w_M = 1/2$. 
This modified scaling behavior leads to a generalized Gibbs--Duhem constraint, encoding the non-trivial relation among the thermodynamic variables that arises from the quasi-homogeneous structure of the fundamental equation. Using the fundamental equation, one obtains the dual thermodynamic variables as  equations of state 
\begin{align}
T
&=
\frac{1-\gamma}{4\sqrt{\pi S}}
+\frac{2P}{\sqrt{\pi}}\sqrt{S}
+\frac{\lambda}{4S}
-\frac{\alpha}{12\sqrt{\pi S}}\,
\frac{
\beta^{2}+3\beta\sqrt{S/\pi}+3S/\pi
}{
(\beta+\sqrt{S/\pi})^{3}
}
+\frac{\alpha}{4\pi}\,
\frac{
\beta^{2}+3\beta\sqrt{S/\pi}+3S/\pi
}{
(\beta+\sqrt{S/\pi})^{4}
}, \label{temp funda}
\\[0.6em]
V&=
\frac{4}{3\sqrt{\pi}}\,S^{3/2},
\label{volume}\\[0.6em]
\Pi_\alpha&=
\frac{
3 \sqrt{\pi}\, S 
+ 3 \pi \sqrt{S}\, \beta 
+ \pi^{3/2} \beta^{2}
}{
6 \left( \sqrt{S} + \sqrt{\pi}\, \beta \right)^{3}
},
\quad
\Pi_\beta=
-\frac{
\pi \alpha \left(
6 S 
+ 4 \sqrt{\pi}\, \sqrt{S}\, \beta 
+ \pi \beta^{2}
\right)
}{
6 \left( \sqrt{S} + \sqrt{\pi}\, \beta \right)^{4}
},
\quad
\Pi_\lambda=
\frac{1}{4}\left(-2+\ln\frac{S}{\pi \lambda^2}\right).
\end{align}
Equivalently, in terms of the horizon radius $r_h=\sqrt{S/\pi}$,
\begin{equation}
T=\frac{1}{4\pi}\,f'(r_h),
\qquad
V=\frac{4\pi}{3}\,r_h^{3}. \label{geo vol}
\end{equation}

\begin{figure}[ht!]
\centering
    \includegraphics[width=0.45\linewidth]{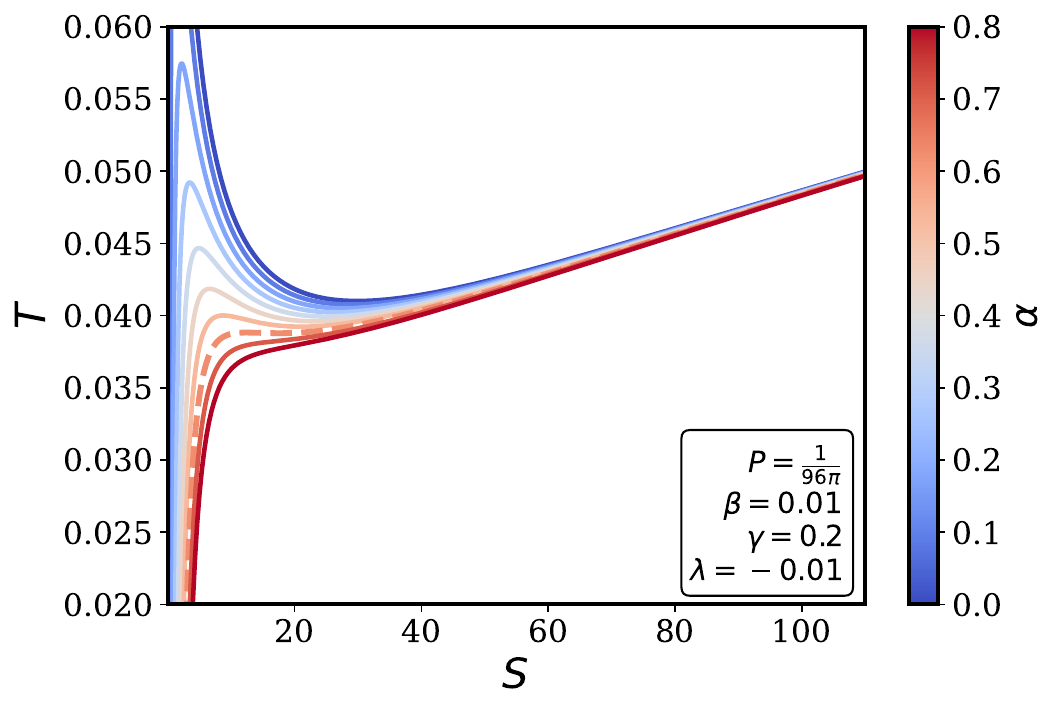}\quad
    \includegraphics[width=0.5\linewidth]{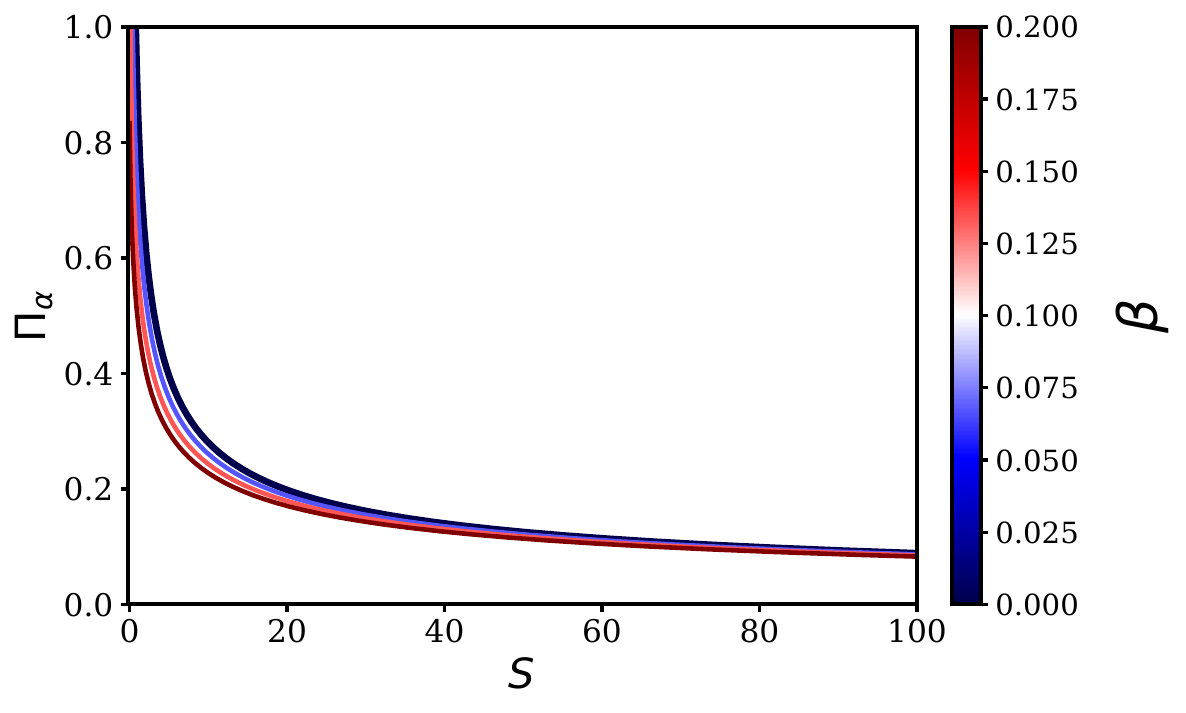}\\
    (i)  \hspace{6cm} (ii)\\
    \includegraphics[width=0.45\linewidth]{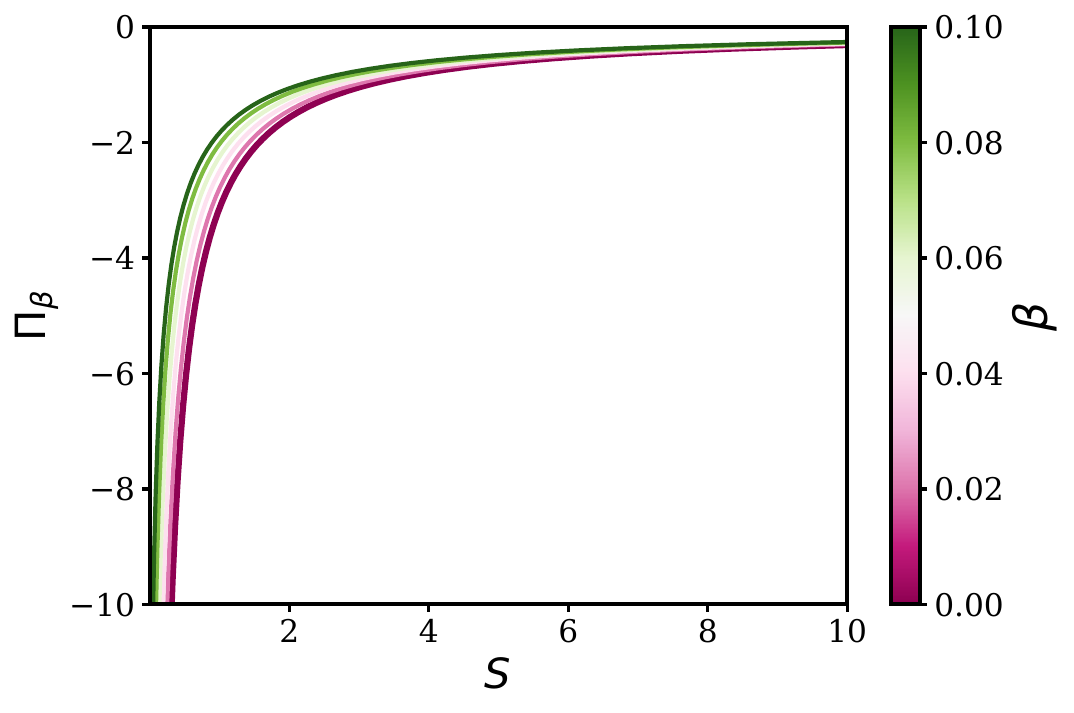}
    \includegraphics[width=0.5\linewidth]{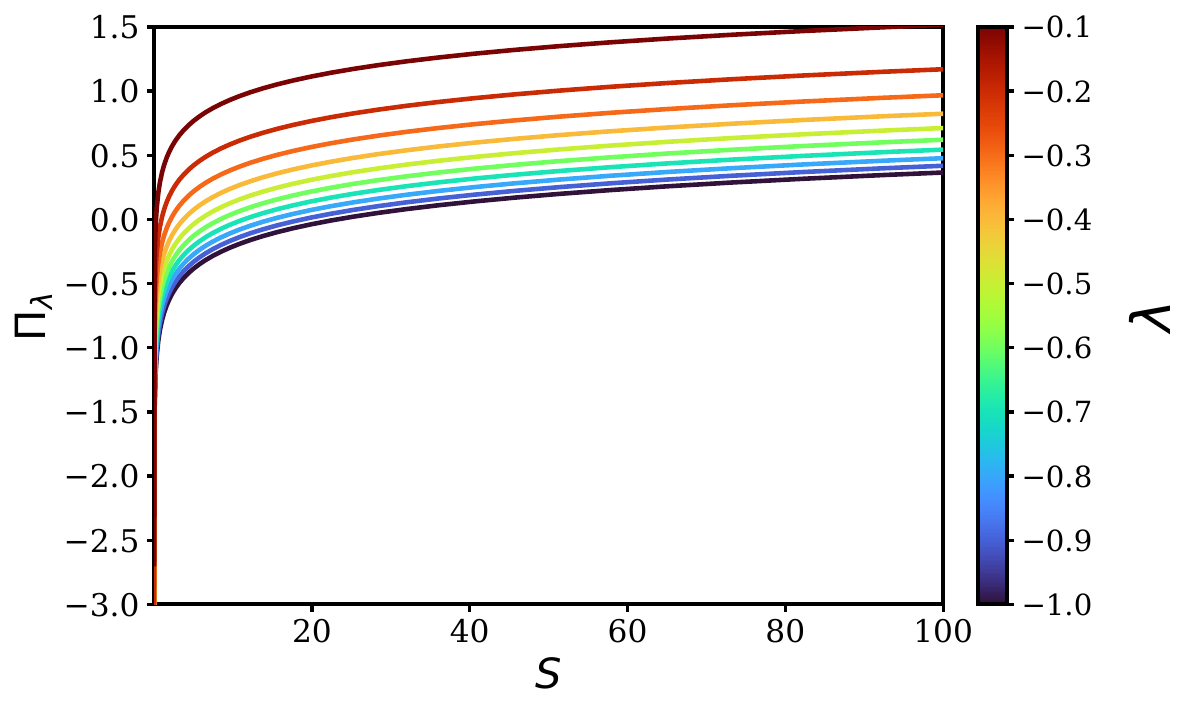}\\
    (iii)  \hspace{6cm} (iv)
\caption{Thermodynamic functions of the deformed AdS black hole: 
(i) temperature $T(S)$, 
(ii) $\Pi_\alpha(S)$, 
(iii) $\Pi_\beta(S)$, and 
(iv) $\Pi_\lambda(S)$, 
with the remaining parameters fixed.}
\label{temperature_function}
\end{figure}

Notice that $\Pi_\alpha$ has the correct scaling of an electric potential (inverse length), supporting its interpretation as a generalized electric-type potential Fig.\ref{temperature_function}(ii) associated with the charge-like parameter $\alpha$. In contrast, $\Pi_\beta$ and $\Pi_\lambda$ are dimensionless and thus  they should be understood as response coefficients. Moreover, despite the presence of a CS, PFDM, and deformation parameters, the thermodynamic volume Eq.~\eqref{volume} coincides with the geometric volume Eq.~\eqref{geo vol}. Consequently, the reverse isoperimetric inequality is saturated, indicating that the black hole is not super-entropic \cite{kubizvnak2017black}. However, in the presence of rotation \cite{ladino2025phase},  and other thermodynamic parameters \cite{romero2024extended1}, the formula for the thermodynamic volume gets more complicated, and in general the thermodynamic volume does not seem to have a geometrical meaning.

\begin{figure}[ht!]
\centering
\includegraphics[width=0.75\linewidth]{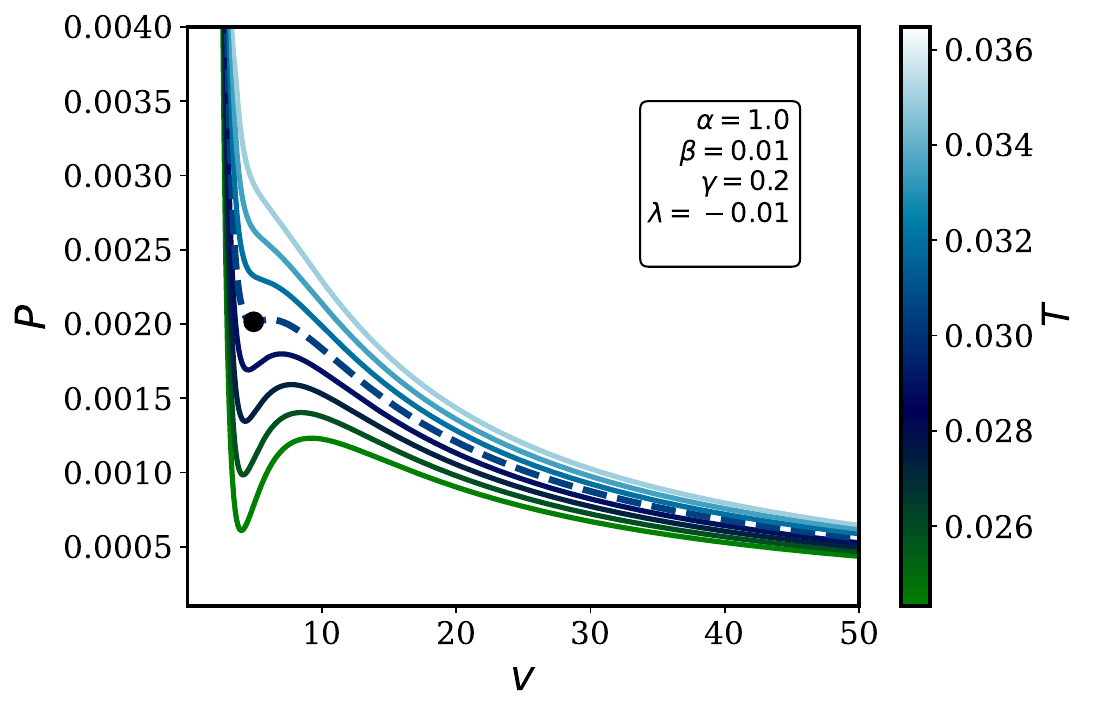}
\caption{
Equation of state $P(v,T)$ for the deformed AdS black hole with
fixed $\alpha,\beta,\gamma,\lambda$.
The dashed curve corresponds to the critical isotherm. The black dot marks the critical point
$(P_c,v_c,T_c) \simeq (0.00201,\,5.337,\,0.0304)$.
}
\label{fig:PV_profile}
\end{figure}

As shown in Fig.~\ref{temperature_function}(ii)--(iv), all conjugate potentials remain regular and monotonic functions of the entropy throughout the physically allowed region. In contrast, the temperature function displays a much richer structure, developing up to two local extrema or, alternatively, an inflection point (dashed curve), as illustrated in Fig.~\ref{temperature_function}(i). We therefore concentrate on how the presence of surrounding matter fields and geometric deformations modifies the thermodynamic behavior relative to the RN-AdS solution. In particular, we analyze how these additional contributions displace the local minimum, local maximum, and the critical (inflection) point of the temperature curve. The temperature develops local extrema when
$\partial T/\partial S= 0$, which leads to
\begin{align}
\frac{\partial T}{\partial S} =\;&
8 P S^{4}
+ 40 P \sqrt{\pi}\, S^{7/2} \beta
+ S^{3} \left(-1 + 80 P \pi \beta^{2} + \gamma \right) + \pi^{3/2} S^{3/2} \beta 
\left[
-\alpha 
+ 2\beta \left(4 P \pi \beta^{3} + 5 \beta (-1 + \gamma) - 10 \lambda \right)
\right]
\nonumber\\
&+ \pi S^{2}
\left[
3 \alpha 
+ 10 \beta \left(\beta (-1 + 4 P \pi \beta^{2} + \gamma) - \lambda \right)
\right]+ \pi^{5/2} S^{1/2}\, \beta^{4}
\left[
\beta (-1 + \gamma) - 10 \lambda
\right] \label{Textrem}\\
&+ 5 \pi^{2} S \beta^{3}
\left[
\beta (-1 + \gamma) - 4 \lambda
\right]+ \sqrt{\pi} S^{5/2}
\left[
80 P \pi \beta^{3} + 5 \beta (-1 + \gamma) - 2 \lambda
\right]
- 2 \pi^{3} \beta^{5} \lambda
= 0 .\nonumber 
\end{align}
For \(\gamma = \beta = \lambda = 0\) the above expression reduces to 
\begin{equation}
8 P S^{2} - S + 3 \pi \alpha = 0.
\end{equation}
whose solutions are given by
\begin{equation}
S_{0}(\alpha,P)=
\frac{1 \pm \Delta}{16 P},\qquad \Delta\equiv \sqrt{1 - 96 \pi \alpha P} . \label{RN S_extr}
\end{equation}
Identifying $\alpha = Q^{2}$, one recovers the standard RN--AdS result \cite{ladino2024phase}. However, in the general case, Eq.~\eqref{Textrem} does not admit a closed-form analytic solution. Therefore, an exact treatment is not feasible, and we instead consider a perturbative approximation in the regime of small $\beta$ and $\lambda$
\begin{equation}
\mathcal{T}_s \equiv \left.
\frac{\partial T}{\partial S}
\right|_{\beta,\lambda \ll 1}
\simeq
\frac{
8 P S^{2}
+ 3 \pi \alpha
+ S(-1+\gamma)
- 16 \sqrt{\pi}\,\alpha \beta\, S^{-1/2}
- 2 \sqrt{\pi}\,\lambda\, S^{1/2}
}{
8 \sqrt{\pi}\, S^{5/2}
}\, .
\label{ts corr}
\end{equation}
For small deformations, we expect the extremal entropy to be slightly shifted with respect to the RN--AdS solution, and thus write $S_{\text{ext}} \simeq S_{0} + \delta S$, where $S_{0}$ is given by Eq.~\eqref{RN S_extr}, at first order we obtain
\begin{equation}
\mathcal{T}_s(S_0+\delta S)\simeq \mathcal{T}_s(S_0)+\mathcal{T}_{ss}(S_0)\delta S=0,
\end{equation}
from which it follows that
\begin{equation}
\delta S=-\frac{\mathcal{T}_s(S_0)}{\mathcal{T}_{ss}(S_0)}.
\end{equation}
Thus, by substituting Eq.~\eqref{ts corr}, the first-order correction to the extremal entropy can be expressed as
\begin{equation}
\delta S
=
\pm\,
\frac{
-\,2\gamma S_0^{3/2}
+4\sqrt{\pi}\lambda S_0
+32\pi^{3/2}\alpha\beta
}{
\Delta\sqrt{S_0}
}.
\end{equation}
The upper and lower signs correspond to the two local extremal points. In the limit $\Delta \to 0$, the extrema coalesce, signaling the onset of a critical point. To determine the corresponding critical quantities explicitly, we now turn to the general equation of state $P(v,T)$ of the deformed AdS black hole, written in terms of the specific volume $v = 2 r_h$  \cite{kubizvnak2017black},
\begin{multline}
P(v,T)=
\frac{1}{2\pi v^3}
\Bigg[
\frac{
v \Big(
2\pi T v^5 
+ 16 \beta^4 (-1 + \gamma)
+ 32 v \beta^3 (-1 + \pi T \beta + \gamma)
+ 8 v^3 \beta (-1 + 6 \pi T \beta + \gamma)
\Big)
}{(v+2\beta)^4}
\\
+
\frac{
v \Big(
v^4 (-1 + 16 \pi T \beta + \gamma)
+ 4 v^2 \left(\alpha + 2 \beta^2 (-3 + 8 \pi T \beta + 3 \gamma)\right)
\Big)
}{(v+2\beta)^4}
- 2\lambda
\Bigg].
\end{multline}
The critical point $(P_c,v_c,T_c)$ is determined by the conditions\footnote{Notation: $\Phi_{,xy\ldots}$ denotes the partial derivative 
$\partial^n \Phi / \partial x \partial y \ldots$, where the subscripts 
indicate differentiation with respect to the corresponding variables.}
\begin{equation}
\mathcal{P}_{,v}\equiv\left(\frac{\partial P}{\partial v}\right)_T = 0,
\qquad
\mathcal{P}_{,vv}\equiv\left(\frac{\partial^2 P}{\partial v^2}\right)_T = 0. \label{PV}
\end{equation}
The simultaneous fulfillment of Eqs.~\eqref{PV} yields the criticality condition
\begin{equation}
\left. P_{,vv} \right|_{T=T_c}= v(\gamma-1) (v+2\beta)^6
+8\alpha v^4(3v-4\beta)
-6\lambda (v+2\beta)^6
=0.\label{PV1}
\end{equation}
The criticality condition gives rise to a seventh-degree polynomial in $v$. For this reason, we adopt a perturbative approach and work in the regime $\beta \ll v$, where
\begin{equation}
(v+2\beta)^{-6}\simeq
v^{-6}
\left(
1-\frac{12\beta}{v}
\right).
\end{equation}
Substituting into the criticality condition Eq. \eqref{PV1} yields
\begin{equation}
\left. P_{,vv} \right|_{T=T_c}\simeq (\gamma-1)v^3
-6\lambda v^2
+24\alpha v
-320\alpha\beta
=0. \label{PVv}
\end{equation}
We expand around the undeformed RN--AdS solution. In this limit Eq.~\eqref{PVv} reduces to
\begin{equation}
- v^3 + 24\alpha v = 0,
\end{equation}
whose nontrivial solution is the RN--AdS critical volume $v_0 = \sqrt{24\alpha}$ \cite{kubizvnak2017black}. At first order, we have
\begin{equation}
v_c = v_0 + \delta v, \qquad  P_{,vv}(v_0+\delta v)
\simeq
 P_{,vv} (v_0)
+
 P_{,vvv}(v_0)\,\delta v,
\end{equation}
the correction is then given by
\begin{equation}
\delta v
=
-\frac{\mathcal{P}_{,vv}(v_0)}{\mathcal{P}_{vvv}(v_0)}.
\end{equation}
Keeping terms linear in $\gamma$, $\lambda$ and $\beta$, one obtains
\begin{equation}
\delta v
=
\frac{\gamma}{2}\,v_0
-\frac{3\lambda}{2}
-\frac{20}{3}\beta.
\end{equation}
Therefore, the corrected critical specific volume becomes
\begin{equation}
v_c
\simeq
\sqrt{24\alpha}\left[1+\frac{\gamma}{2}
-\frac{1}{\sqrt{24\alpha}}\left(\frac{3\lambda}{2}
+\frac{20}{3}\beta\right)\right].\label{vc correc}
\end{equation}
Following the same reasoning, and using the corrected critical volume 
given in Eq.~\eqref{vc correc}, we evaluate the corresponding corrections 
to the critical temperature and critical pressure, which read
\begin{align}
T_c &\simeq
\frac{\sqrt{6}}{18\pi\sqrt{\alpha}}
\left(
1-
\frac{3}{2}\gamma
+
\frac{10\beta+9\lambda}{4\sqrt{6}\sqrt{\alpha}}
\right), \label{tc correc}
\\[6pt]
P_c &\simeq
\frac{1}{96\pi\alpha}
\left(
1
-2\gamma
+
\frac{2\sqrt{6}(4\beta+3\lambda)}{9\sqrt{\alpha}}
\right). \label{pc correc}
\end{align}
As a result, using Eqs.~\eqref{vc correc}-- \eqref{pc correc}, 
the universal RN--AdS ratio $P_c v_c/T_c$ can be expressed in the compact form
\begin{equation}
\frac{P_c v_c}{T_c}
\simeq
\frac{3}{8}
\left(
1
-
\frac{\sqrt{6}\,(\beta - 2\lambda)}{12\sqrt{\alpha}}
\right).
\end{equation}
Remarkably, no $\gamma$ dependence appears at first order, indicating that the CS does not modify this thermodynamic universal constant at the linear level.  Similar modifications of the critical ratio are known to occur in higher-dimensional black holes and in alternative or modified gravity theories \cite{gunasekaran2012extended}, where additional couplings and geometric contributions alter the effective equation of state. Fig.~\ref{fig:PV_profile} shows the $P$--$V$ diagram for the deformed black hole solution with chosen parameter values. 

\subsection{Local Thermodynamic Stability}

According to classical thermodynamics, the local stability and phase
transition structure of the system are determined by the behavior of the generalized heat capacities~\cite{romero2024extended1}. Within a general framework, the Nambu bracket formalism
(see the Appendix of~\cite{ladino2024phase})
provides a coordinate-independent definition of the heat capacity
at fixed thermodynamic parameters $(x^1,x^2,\ldots,x^{n-1})$
in the equilibrium state space
$(y^1,y^2,\ldots,y^{n})$
\begin{align}
C_{x^1,\ldots,x^{n-1}}
=
T\left(\frac{\partial S}{\partial T}\right)_{x^1,\ldots,x^{n-1}}
=
T\,
\frac{\{S,x^1,\ldots,x^{n-1}\}_{y^1,\ldots,y^{n}}}
{\{T,x^1,\ldots,x^{n-1}\}_{y^1,\ldots,y^{n}}}.
\label{nambu1}
\end{align}
For the deformed AdS--black hole, in the thermodynamic representation
determined by the first law Eq.~\eqref{first law},
the relevant response function is the heat capacity computed
at constant $X \equiv \{P,\alpha,\beta,\lambda\}$,
which is given by
\begin{align}
C_{X}&=T\left(\frac{\partial S}{\partial T}\right)_{X}=\frac{T}{M_{,SS}}=\frac{\mathcal{N}}{\,\mathcal{D}}, \label{heat capacityx}
\end{align}
where the numerator and denominator are
\begin{equation}
\begin{aligned}
\mathcal{N}
&=
-2S \left(\sqrt{S}+\sqrt{\pi}\beta\right)
\Bigg[
\lambda \pi^{3/2}\left(\sqrt{S}+\sqrt{\pi}\beta\right)^4+ 8\pi PS
\\
&\quad
+\pi \sqrt{S}
\Big(
S^2
-\pi S\alpha
+4\sqrt{\pi}S^{3/2}\beta
+6\pi S\beta^2
+4\pi^{3/2}\sqrt{S}\beta^3
+\pi^2\beta^4
-\left(\sqrt{S}+\sqrt{\pi}\beta\right)^4\gamma
\Big)
\Bigg],\\[0.3em]
\mathcal{D}
&=\;
2\pi^{4}\beta^{5}\lambda
+5\pi^{3}S\beta^{3}\left(\beta-\beta\gamma+4\lambda\right)
+\pi^{7/2}\sqrt{S}\,\beta^{4}\left(\beta-\beta\gamma+10\lambda\right)
\\[0.3em]
&-8\pi P S^{4}
-40\pi^{3/2} P S^{7/2}\beta
+\pi S^{3}\left(1-\gamma-80\pi P \beta^{2}\right)
\\[0.3em]
&+\pi^{3/2}S^{5/2}
\left[
-5\beta(-1+\gamma)
+2\lambda
-80\pi P \beta^{3}
\right]
\\[0.3em]
&+\pi^{2}S^{2}
\left[
-3\alpha
+5\beta
\left(
2\beta-2\beta\gamma+2\lambda
-8\pi P \beta^{3}
\right)
\right]
\\[0.3em]
&+\pi^{5/2}S^{3/2}\beta
\left[
\alpha
+\beta
\left(
-10\beta(-1+\gamma)
+20\lambda
-8\pi P \beta^{3}
\right)
\right].
\end{aligned}
\end{equation}

\begin{figure}[ht!]
\centering
\begin{minipage}{0.49\textwidth}
\centering
\includegraphics[width=\linewidth]{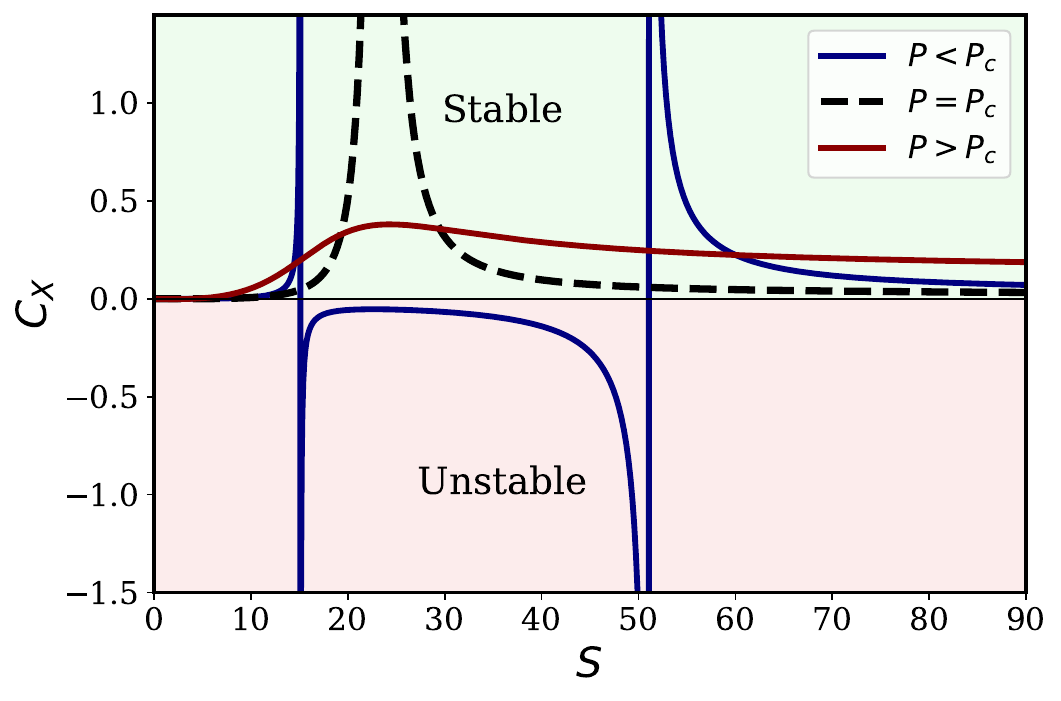}
\end{minipage}
\hfill
\begin{minipage}{0.49\textwidth}
\centering
\includegraphics[width=\linewidth]{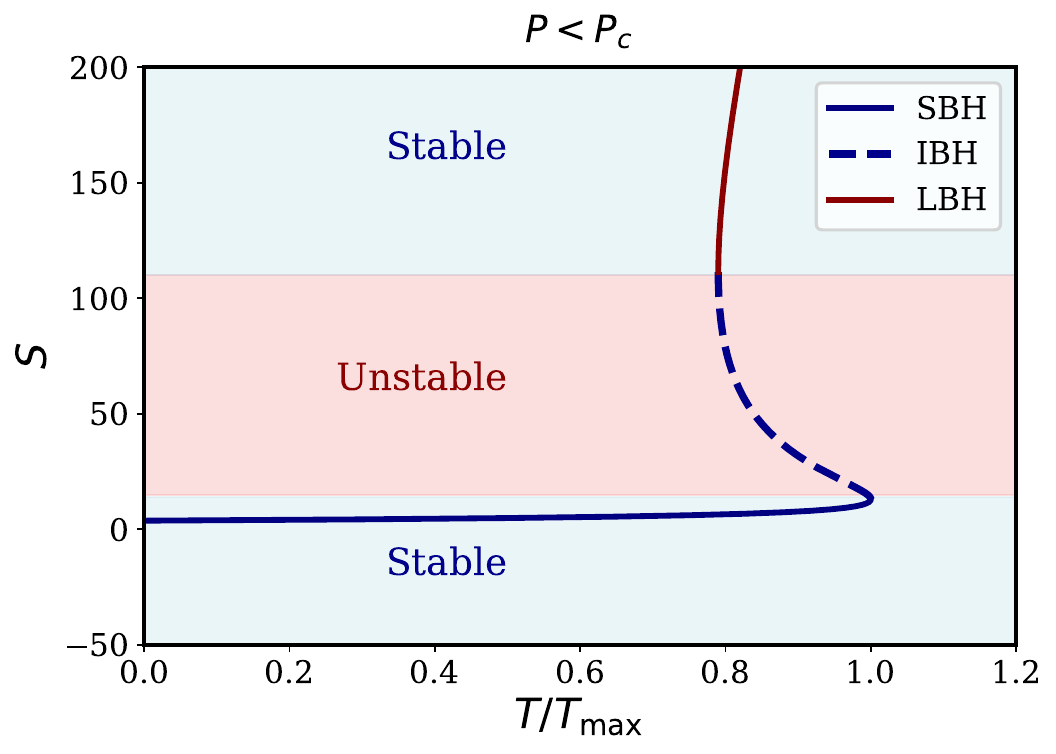}
\end{minipage}
\caption{
Left: Heat capacity of the deformed AdS black hole at 
$\alpha=1$, $\beta=0.01$, $\gamma=0.2$, $\lambda=-0.01$. 
The dashed curve denotes $C_X$ at critical pressure $P_c \simeq 0.00201$, 
while solid curves correspond to $P<P_c$ and $P>P_c$. 
Right: Normalized $T/T_{\max}$ diagram in the subcritical regime, 
displaying the small (SBH), intermediate (IBH), and large (LBH) 
black hole branches.
}
\label{fig:heat capacity}
\end{figure}

As shown in Fig.~\ref{fig:heat capacity}, the heat capacity displays the
characteristic behavior of a thermodynamic system with a single critical point.
For $P<P_c$, two divergences appear, delimiting an intermediate unstable
branch where $C_X<0$ and signaling a first–order phase transition between
the small and large black hole phases. As the pressure increases to the
critical value $P=P_c$, the two singularities coalesce into a single
divergence, marking the onset of a second–order phase transition.
For $P>P_c$, the heat capacity becomes continuous and finite,
indicating the disappearance of the phase transition and the emergence
of a supercritical regime.

\subsection{Phase Structure}

\begin{figure}[ht!]
\centering
\begin{minipage}{0.49\textwidth}
    \centering
    \includegraphics[width=\linewidth]{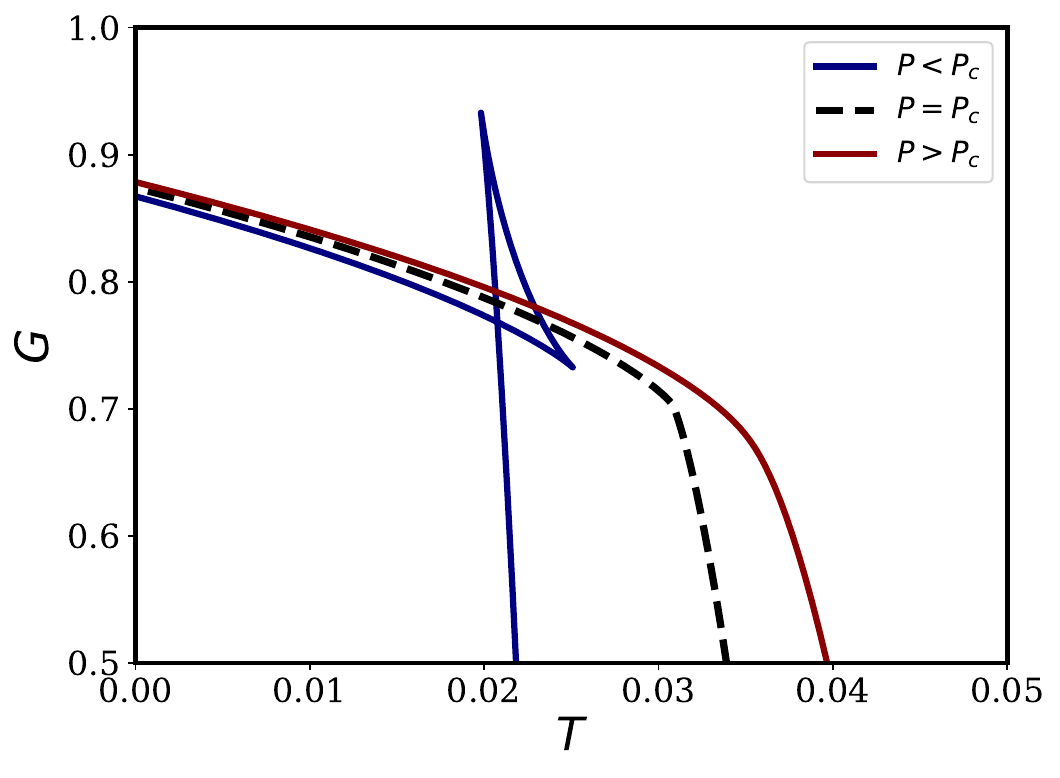}
\end{minipage}
\hfill
\begin{minipage}{0.49\textwidth}
    \centering
    \includegraphics[width=\linewidth]{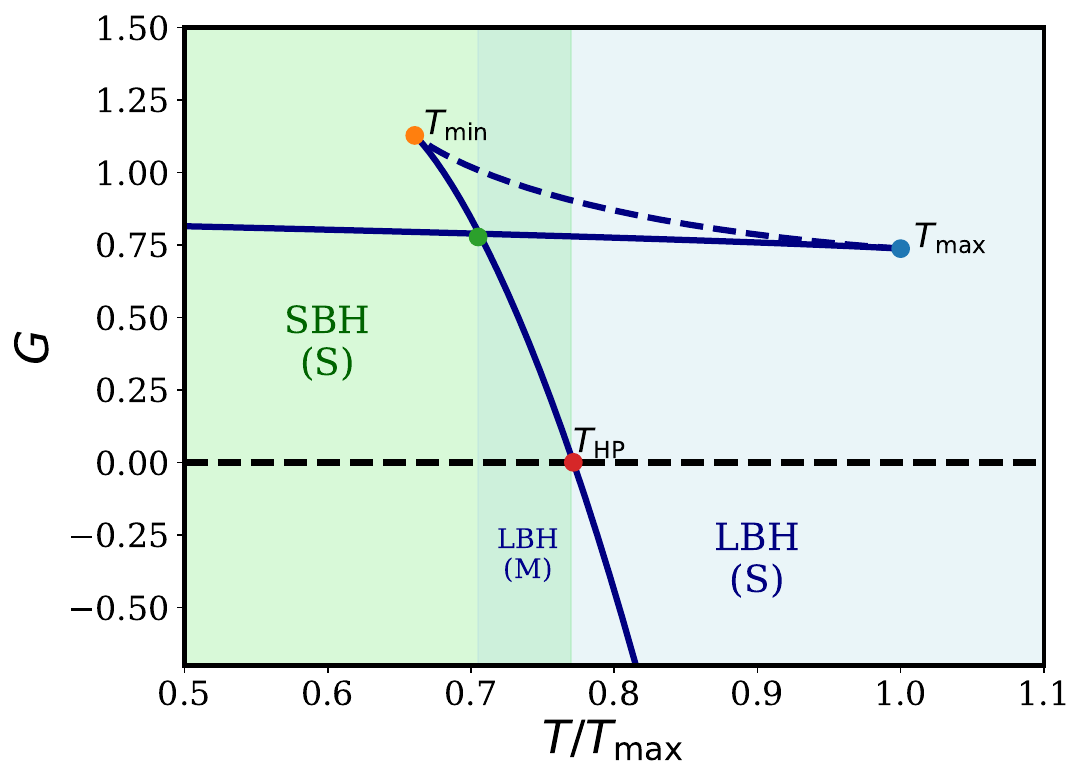}
\end{minipage}
\caption{
Gibbs free energy for the deformed AdS black hole with 
$\alpha=1$, $\beta=0.01$, $\gamma=0.2$, and $\lambda=-0.01$. 
(a) For $P>P_c$, the system exhibits a single thermodynamic phase; 
for $P<P_c$, three branches appear. 
At $P=P_c \simeq 0.0021$, the curves merge, signaling a second-order phase transition. 
(b) Case $P<P_c$ with $P=0.0005$, displaying the SBH, IBH (dashed blue curve), and LBH branches. 
Stable (S) and metastable (M) regions are indicated accordingly. The green marker indicates $T_{\mathrm{coex}}$, where the first-order phase transition occurs.
}
\label{fig:free_energy}
\end{figure}

To characterize the global thermodynamic stability and phase structure of the deformed--AdS black hole, we analyze the Gibbs free energy in the extended phase space, where the mass is interpreted as gravitational enthalpy \cite{kubizvnak2017black}. This potential enables the identification of phase transitions and determines the thermodynamically preferred branches. Accordingly, the Gibbs free energy is defined as $G \equiv M - TS$
\begin{equation}
   G(S,P)=
\frac{1}{12 \sqrt{\pi}\left(\sqrt{S}+\sqrt{\pi}\,\beta\right)^4}
\Bigg[G^{Schw}+G^{def}
\Bigg], \label{free energy eq} 
\end{equation}
where $G^{Schw}$ denotes the Schwarzschild–AdS contribution \cite{ladino2024phase}, and $G^{def}$ contains the PFDM, CoS and deformed sector
\begin{align}
G^{Schw} &\equiv
3 S^{5/2}
- 8 P S^{7/2},
\label{G_RN}
\\[8pt]
G^{def} &\equiv
12 \sqrt{\pi} S^2 \beta
- 32 P \sqrt{\pi} S^3 \beta
+ 18 \pi S^{3/2} \beta^2
- 48 P \pi S^{5/2} \beta^2
\nonumber\\
&\quad
+ 12 \pi^{3/2} S \beta^3
- 32 P \pi^{3/2} S^2 \beta^3
+ 3 \pi^2 \sqrt{S} \beta^4
- 8 P \pi^2 S^{3/2} \beta^4
\nonumber\\
&\quad  
+ \alpha \Big(9 \pi S^{3/2}+
12 \pi^{3/2} S \beta
+ 8 \pi^2 \sqrt{S} \beta^2
+ 2 \pi^{5/2} \beta^3
\Big)
\nonumber\\
&\quad
- \gamma \Big(
3 S^{5/2}
+ 12 \sqrt{\pi} S^2 \beta
+ 18 \pi S^{3/2} \beta^2
+ 12 \pi^{3/2} S \beta^3
+ 3 \pi^2 \sqrt{S} \beta^4
\Big)
\nonumber\\
&\quad
+ \lambda \Big(
- 3 \sqrt{\pi} S^2
- 12 \pi S^{3/2} \beta
- 18 \pi^{3/2} S \beta^2
- 12 \pi^2 \sqrt{S} \beta^3
- 3 \pi^{5/2} \beta^4
\Big)
\nonumber\\
&\quad
+ \lambda \ln\!\left(
\frac{\sqrt{S}}{\sqrt{\pi}|\lambda|}
\right) \Big(
6 \sqrt{\pi} S^2
+ 24 \pi S^{3/2} \beta
+ 36 \pi^{3/2} S \beta^2
+ 24 \pi^2 \sqrt{S} \beta^3
+ 6 \pi^{5/2} \beta^4
\Big)
.
\label{G_def}
\end{align}
From Eq.~(\ref{temp funda}) it follows that $S$ is implicitly determined by the thermodynamic variables. As a consequence, the thermodynamic branches cannot be obtained in closed analytic form. Nevertheless, employing the linear approximation around the critical point given in Eqs.~\eqref{vc correc}--\eqref{pc correc}, we plot the free energy in Fig.~\ref{fig:free_energy}. 

For $P<P_c$, the system exhibits three thermodynamic phases, namely the 
SBH, IBH,and LBH phases, characterized by the typical swallow-tail behavior of the Gibbs free energy. This structure signals the presence of a coexistence curve and a first-order phase transition between the SBH and LBH phases. 
At $P=P_c$, the swallow-tail structure collapses to a single point, the coexistence curve vanishes, and the phase transition becomes second order. For $P>P_c$, the system enters a supercritical regime with a single stable phase. Overall, the numerical analysis indicates that the deformed AdS black hole retains the VdW behavior of the RN--AdS case. However, when the deformation parameters and matter couplings become sufficiently large, driving the system away from the RN--AdS limit, the phase transition structure disappears. Furthermore, Fig.~\ref{fig:free_energy}(b) illustrates the case $P<P_c$, 
showing all possible thermodynamic states, including the global thermal AdS background ($G=0$). At low temperatures, the thermal AdS phase is globally preferred, 
so radiation dominates the spacetime \cite{aa19}. 
As the temperature increases, a LBH can nucleate, with the Hawking-Page (HP) phase transition occurring at $T_{HP}$. 
For $T > T_{HP}$, the stable-LBH becomes the globally preferred configuration.Next, we analyze the effect of the deformation parameters and surrounding fields on the HP transition.

\begin{center}
{\bf Effects of CS and PFDM on the Hawking-Page Transition}
\end{center}

\begin{figure}[ht!]
\centering

\begin{minipage}{0.48\textwidth}
    \centering
    \includegraphics[width=\linewidth]{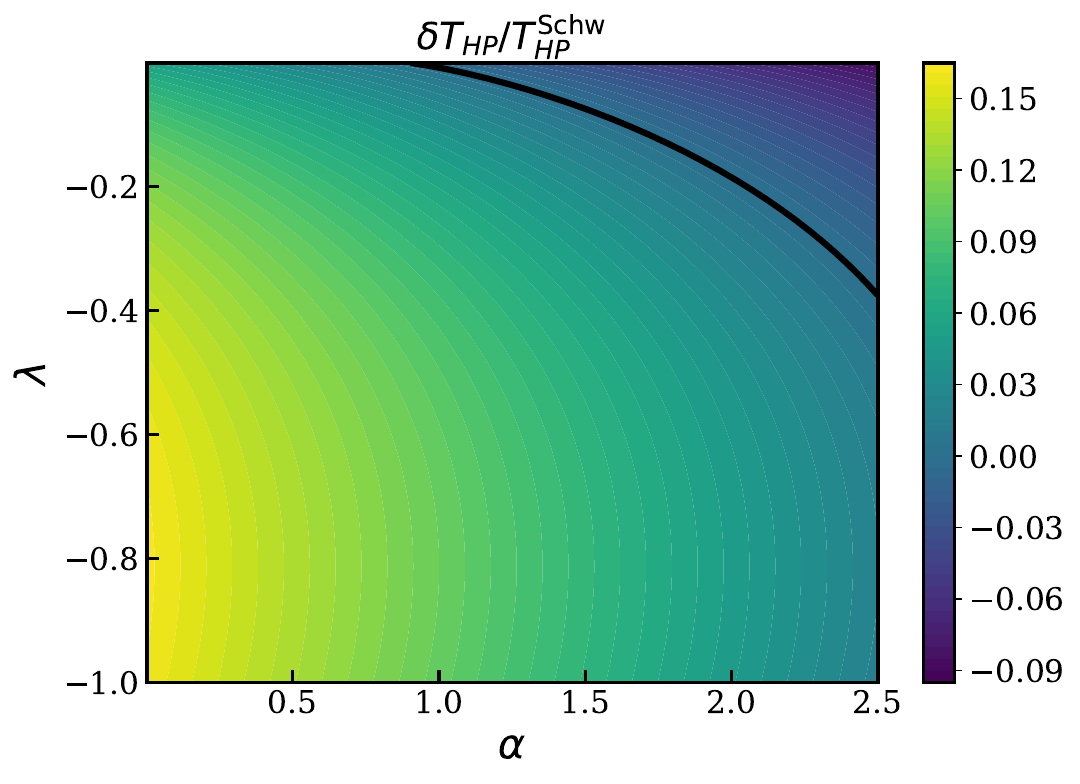}
    \\
    \small (a)
\end{minipage}
\hfill
\begin{minipage}{0.48\textwidth}
    \centering
    \includegraphics[width=\linewidth]{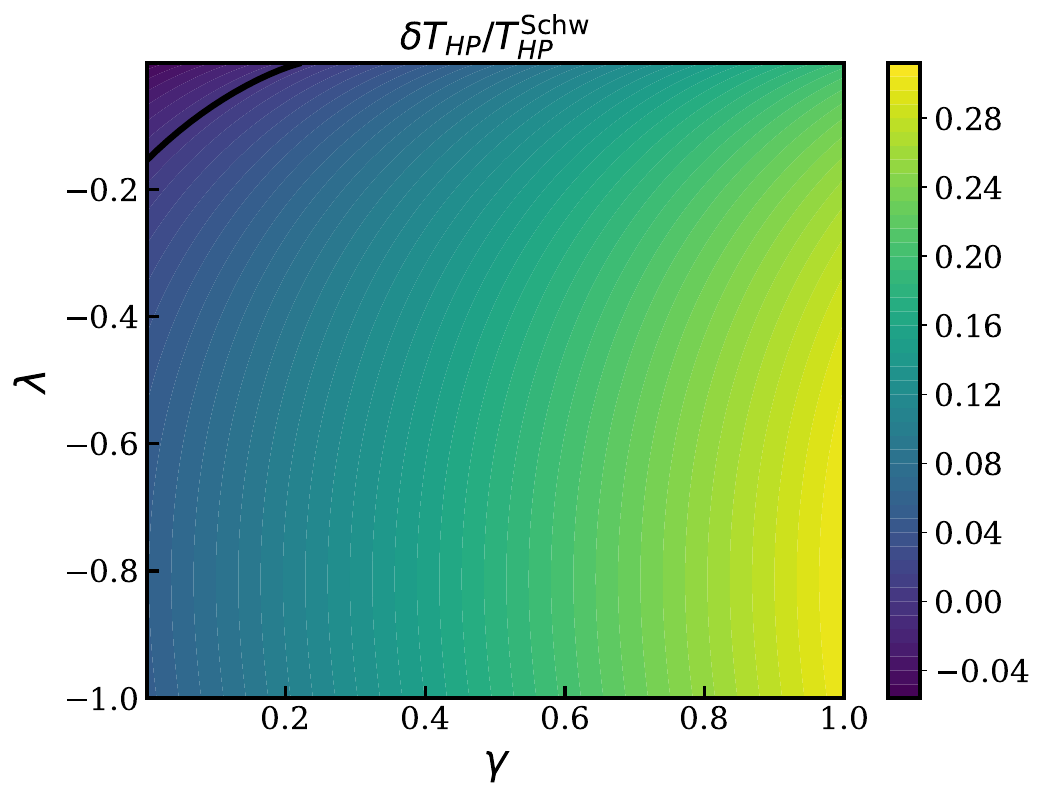}
    \\
    \small (b)
\end{minipage}

\caption{
Contour plots of the relative shift of the HP temperature,
$\delta T_{HP}/T_{HP}^{\mathrm{Schw}}$, for $P=0.009$.
(a) Dependence on the deformation parameter $\alpha$ and the PFDM parameter $\lambda$ with $\gamma=0.2$ fixed.
(b) Dependence on the cloud-of-strings parameter $\gamma$ and $\lambda$ with $\alpha=1.0$ fixed.
The black solid curve represents the critical locus $\delta T_{HP}=0$, separating regions where $T_{HP}$ increases ($\delta T_{HP}>0$) from those where it decreases ($\delta T_{HP}<0$).
}

\label{fig:HP_shift_contours}
\end{figure}

The HP transition is sensitive to modifications of the gravitational background. For example,  non-relativistic geometries with anisotropic Lifshitz scaling modify the $T_{HP}$ through a dynamical exponent \cite{herrera2023anisotropic,herrera2021hyperscaling}. Similar deviations also appear in higher-dimensional and modified gravity theories.  We examine how the PFDM and the CSS modify the HP temperature. To this end, we analyze the Gibbs 
free energy given in Eq.~(\ref{free energy eq}) and determine its zero, 
which defines the HP transition. Since this condition cannot be solved 
analytically, we follow the strategy of the previous section and compute 
the first-order correction to $T_{HP}$. The HP transition is 
defined by
\begin{equation}
G(S;\beta,\gamma,\lambda)=0.
\end{equation}
Let $S_{HP}$ denote the unperturbed HP solution satisfying
\begin{equation}
G^{Schw}(S_{HP})=0,\qquad S_{HP} = \frac{3}{8 P}.
\end{equation}
Expanding to first order, we obtain
\begin{equation}
\delta S_{HP}
=
-
\frac{ G^{def}(S_{HP})}
{G^{Schw}_{,S}(S_{HP})}.
\end{equation}
For small parameters $\alpha,\beta,\gamma,\lambda \ll 1$ 
and retaining only first order terms in the total perturbative counting, 
we obtain
\begin{equation}
\delta S_{HP} \simeq
-3\pi\alpha
+
\frac{3\gamma}{8P}
+
\frac{\sqrt{6\pi}}{4\sqrt{P}}
\,\lambda
\left(
1+\log\!\left(\frac{8}{3}P\pi\lambda^2\right)
\right)
+
\mathcal{O}(\alpha\beta,\beta^2,\beta\lambda).
\end{equation}
Notice that no linear $\beta$ contribution appears. 
Requiring positive energy density for the CS and PFDM implies 
$\gamma>0$ and $\lambda<0$, respectively. 
In the small-$|\lambda|$ regime, the combination 
$\lambda\!\left(1+\log\!\left(\frac{8}{3}P\pi\lambda^2\right)\right)$ 
is positive for $\lambda<0$, so both $\gamma$ and $\lambda$ 
contribute positively to $\delta S_{HP}$ at linear order. 
Consequently, they shift the HP transition to larger entropy, whereas only the $\alpha$ term produces the opposite effect. Moreover, the corresponding correction to the Schw-AdS HP temperature,
$
T^{\text{Schw}}_{HP}=\sqrt{8P}/\sqrt{3\pi}$, is then given by
\begin{equation}
\begin{aligned}
\delta T_{HP} 
&\simeq
T^{Schw}_{,S}(S_{HP})\, \delta S_{HP}
\\
&=
-4 P^{3/2} \sqrt{\frac{2\pi}{3}}\, \alpha
+
\frac{\sqrt{P}\,\gamma}{\sqrt{6\pi}}+
\frac{2}{3} P \lambda 
\left[
1
+
\log\!\left(\frac{8}{3} P \pi \lambda^2\right)
\right].
\end{aligned}
\end{equation}
According to the results, the CoS and PFDM contributions increase the value of the HP temperature, whereas the parameter $\alpha$ decreases it, as shown in Fig.~\ref{fig:HP_shift_contours}. Consequently, the overall shift in $T_{HP}$ is determined by the competition between the enhancing effects of $\gamma$ and $\lambda$ and the suppressing contribution of $\alpha$.

\section{Probing Black Hole Microstructure through GTD}
\begin{figure}[H]
\centering
\begin{minipage}{0.44\textwidth}
    \centering
    \includegraphics[width=\linewidth]{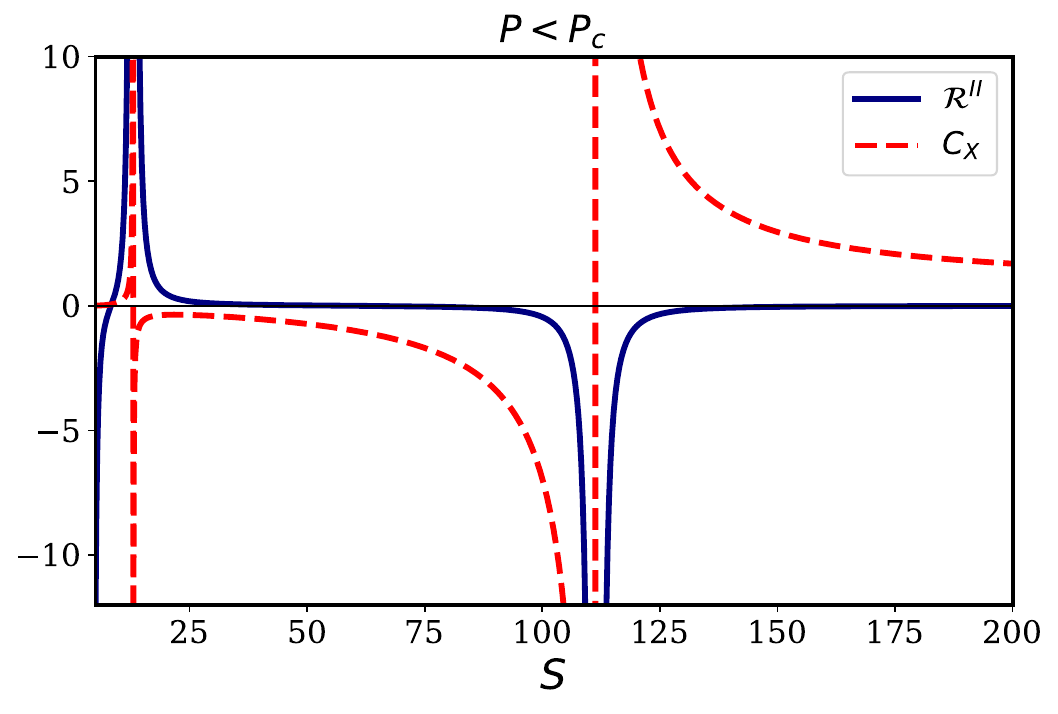}
\end{minipage}
\hfill
\begin{minipage}{0.44\textwidth}
    \centering
    \includegraphics[width=\linewidth]{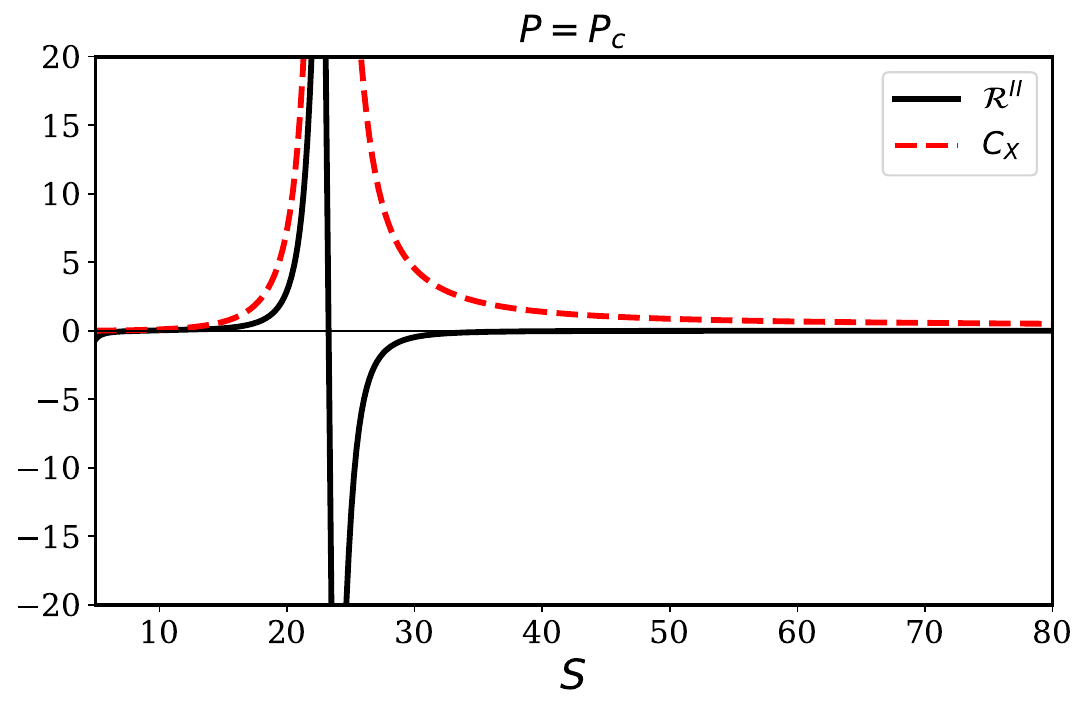}
\end{minipage}
\hfill\\
\begin{minipage}{0.44\textwidth}
    \centering
    \includegraphics[width=\linewidth]{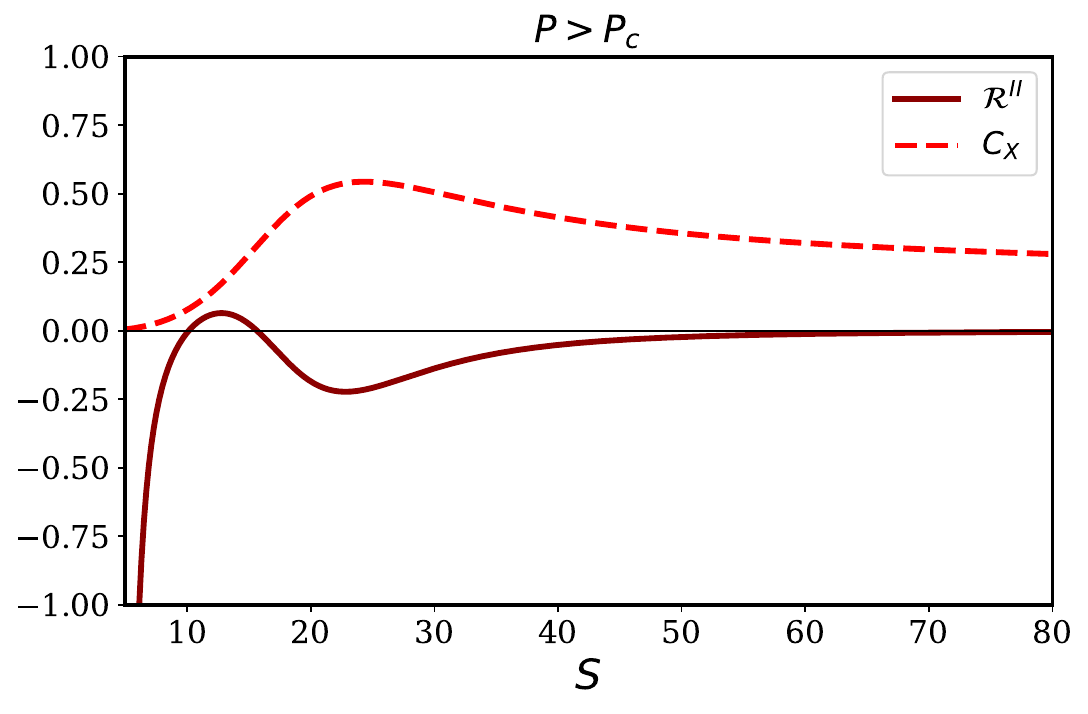}
\end{minipage}

\caption{
Two-dimensional normalized GTD scalar curvature $\mathcal{R}^{II}$ and heat capacity $C_X$ 
as functions of the entropy $S$ for fixed $\alpha=1$, $\beta=0.01$, 
$\gamma=0.2$, and $\lambda=-0.01$. 
The three panels correspond to the subcritical ($P<P_c$), critical ($P=P_c$), 
and supercritical ($P>P_c$) regimes, with $P_c\simeq 0.0021$.
}
\label{fig:GTD scalars}
\end{figure}

In this section, we employ the geometrothermodynamics (GTD) formalism to analyze the thermodynamic geometry of the deformed AdS--black hole, and we examine the correspondence between geometric singularities and the thermodynamic phase structure of the system. GTD provides a geometric framework to investigate the thermodynamic behavior of black holes in a Legendre-invariant manner \cite{quevedo2007geometrothermodynamics}. In the context of ordinary thermodynamics, this invariance ensures that the physical properties of the system remain independent of the choice of thermodynamic potential used for its description. Within this formalism, thermodynamic interactions are encoded in the curvature of the equilibrium manifold, allowing for a direct interpretation of phase transitions and critical phenomena in geometric terms \cite{quevedo2023unified}. In particular, curvature singularities of the GTD scalar are expected to coincide with divergences of thermodynamic response functions, such as the heat capacity, thereby signaling the onset of thermodynamic instabilities and/or phase transitions. Consequently, the analysis of the thermodynamic curvature provides a powerful geometric criterion to investigate the underlying microstructure of black holes \cite{ladino2024phase, ladino2025phase}, and more generally of gravitational systems endowed with horizons \cite{romero2026quasi}. In this framework, the behavior of the curvature scalar encodes information about the effective interactions among the microscopic degrees of freedom.  A positively curved equilibrium space ($\mathcal{R}>0$) is associated with repulsive thermodynamic interactions, whereas a negative curvature ($\mathcal{R}<0$) indicates attractive thermodynamic interactions. In contrast, zero curvature ($\mathcal{R}=0$) corresponds to the absence of interactions, as in the case of an ideal gas, which is geometrically represented by a flat thermodynamic manifold~\cite{quevedo2007geometrothermodynamics}. In GTD, thermodynamic states are represented as points in an $n$-dimensional manifold known as the equilibrium space. In general, there exist three Legendre-invariant metrics in the GTD framework. As shown in \cite{quevedo2023unified}, these metrics are fully compatible and provide equivalent geometric descriptions of the same thermodynamic system. For simplicity, we restrict our analysis to the metric $g^{II}$ in order to probe the thermodynamic microstructure of the black hole. The metric $g^{II}$ is given by
\begin{align}
g^{II} &= \sum_{a,b,c,d=1}^{n} 
\left( \nu_c E^c \frac{\partial \Phi}{\partial E^c} \right) 
\eta^{d}{}_{a} 
\frac{\partial^2 \Phi}{\partial E^b \partial E^d} \, 
dE^a \, dE^b.
\label{g222}
\end{align}
Here, $\Phi$ denotes the thermodynamic potential, $E^c$ are the coordinates of the equilibrium space, $\eta = \mathrm{diag}(-1,1,1,\dots)$ is a constant metric, and $\nu_c$ are, in general, free parameters. However, in order to describe quasi-homogeneous systems, the constants $\nu_c$ are chosen to coincide with the quasi-homogeneity weights $w_i$ of the system, thereby ensuring the validity of the generalized Euler identity (for details, see \cite{quevedo2023unified}). In general, the deformed AdS--black hole, regarded as a quasi-homogeneous system, is described by a five-dimensional equilibrium space with coordinates $E^c = \{S, P, \alpha, \beta, \lambda\}$. However, it can be shown that the corresponding five-dimensional metric Eq.~\eqref{g222} becomes degenerate, since the fundamental Eq.~\eqref{funda} is linear in $P$ and $\alpha$. Consequently, the Hessian with respect to the extensive variables is not of full rank, leading to a vanishing determinant of the induced metric. There are several ways to circumvent this issue. One possible approach is to replace $P$ with the AdS curvature radius $\ell$ as a thermodynamic variable, since $P \propto \ell^{-2}$, this is the aprroach used in \cite{ladino2024phase,ladino2025phase}. In this parametrization, the fundamental equation is no longer linear in the corresponding variable, which removes the degeneracy of the metric. Alternatively, one may consider a reduced equilibrium space associated with isenthalpic processes. For simplicity, we take $S$ and the PFDM parameter $\lambda$ as fluctuation variables. We expect the qualitative results to remain unchanged in higher-dimensional equilibrium spaces, reflecting the universal thermodynamic behavior of gravitational horizons. Indeed, as shown numerically in \cite{romero2026quasi}, the behavior near criticality is independent of the dimension of the equilibrium space. Thus, taking $\{S,\lambda\}$ as independent variables and using the weights given in Eq.~\eqref{wi}, together with the equation of state in Eq.~\eqref{EoS}, the line element in Eq.~\eqref{g222} reduces to
\begin{align}
g^{II}= 
\Sigma
\left(
- M_{,SS} \, dS^2
+ M_{,\lambda\lambda} \, d\lambda^2
\right),
\label{g222_reduced}
\end{align}
where the conformal factor is $\Sigma\equiv TS +\frac{1}{2} \lambda \Pi_\lambda$. The corresponding scalar curvature reads
\begin{align}
\mathcal{R}^{II}
&=
\frac{\mathcal{N}}
{2\,\Sigma^{3}\,M_{,SS}^{2}\,M_{,\lambda\lambda}^{2}},\label{scalar}
\end{align}
where $\mathcal{N}$ is a function of the thermodynamic parameters whose explicit expression is too lengthy to be written here. Notice that in Eq.~\eqref{g222_reduced} the conformal factor does not correspond to the Smarr relation, Eq.~\eqref{smarr}, since we are not working with the complete equilibrium space. Therefore, one expects the appearance of unphysical singularities in Eq.~\eqref{scalar} associated with the zeros of the conformal factor $ \Sigma$. To avoid the spurious singularities arising from this factor, we normalize the scalar curvature. This issue was already solved in \cite{quevedo2023unified}, where it was shown that, for a fully quasi-homogeneous system, there exists a one-to-one correspondence between geometric singularities and the physical divergences associated with phase transitions. Furthermore, observe that the GTD scalar curvature diverges when 
$M_{,SS}=0$, which exactly coincides with the divergences of the heat capacity, 
as depicted in Fig.~\ref{fig:GTD scalars} and given in Eq.~\eqref{heat capacityx}. 
On the other hand, since $M_{,\lambda\lambda} = -\frac{1}{2\lambda}$, 
this term does not generate additional physical singularities.

Therefore, the divergences of the normalized scalar curvature $\mathcal{R}^{II}$ accurately reflect the thermodynamic critical behavior: two distinct singularities arise in the subcritical regime (Fig.~\ref{fig:GTD scalars}(a)), merge into a single critical divergence at $P=P_c$ (Fig.~\ref{fig:GTD scalars}(b)), and disappear in the supercritical region where no phase transition occurs (Fig.~\ref{fig:GTD scalars}(c)). 

\begin{figure}[H]
\centering
\begin{minipage}{0.44\textwidth}
    \centering
    \includegraphics[width=\linewidth]{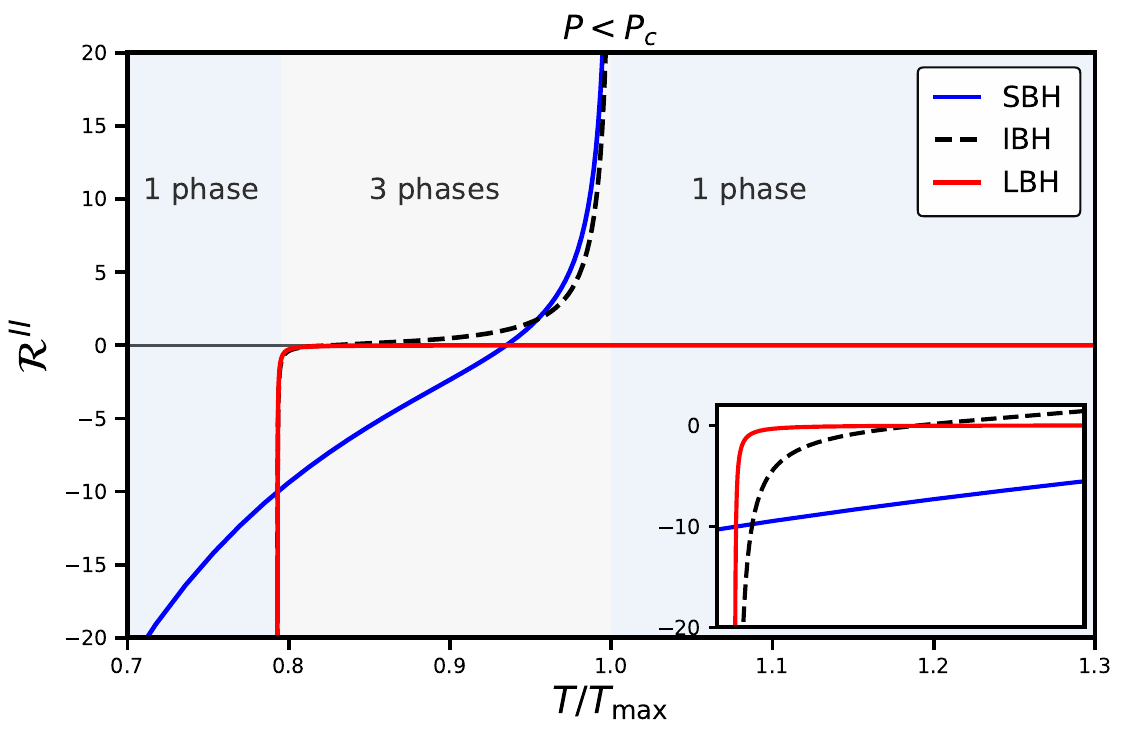}
\end{minipage}
\hfill
\begin{minipage}{0.45\textwidth}
    \centering
    \includegraphics[width=\linewidth]{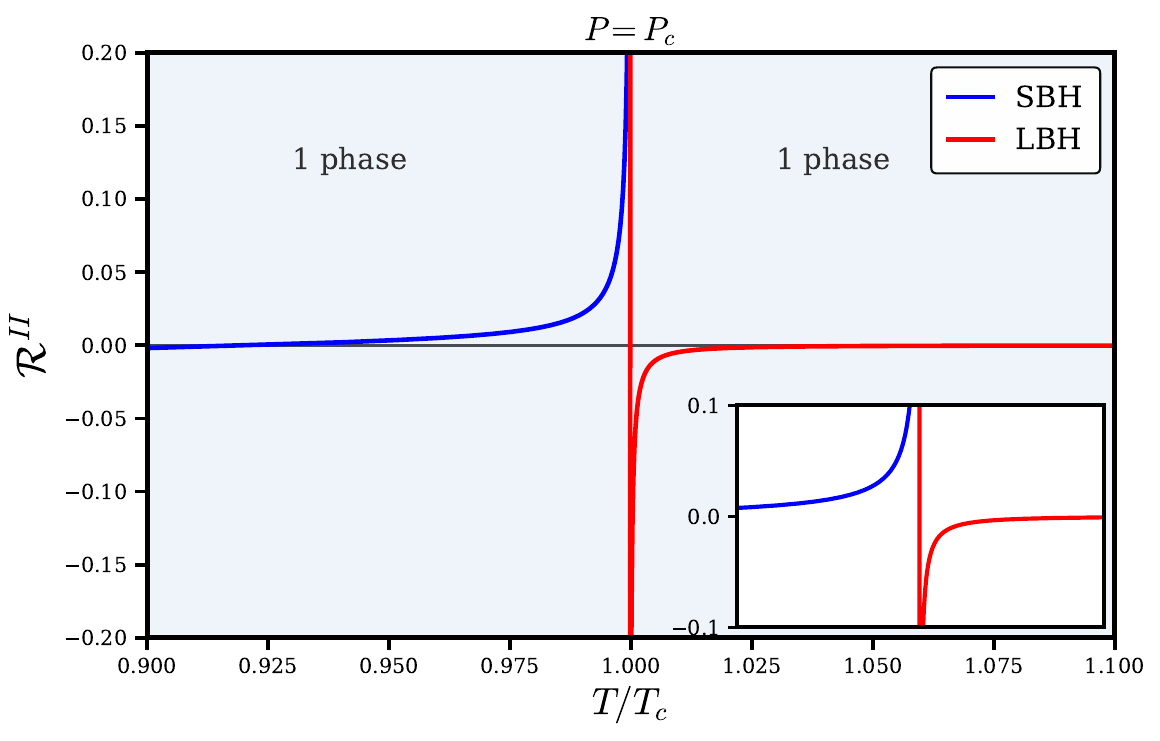}
\end{minipage}
\hfill\\
\begin{minipage}{0.45\textwidth}
    \centering
    \includegraphics[width=\linewidth]{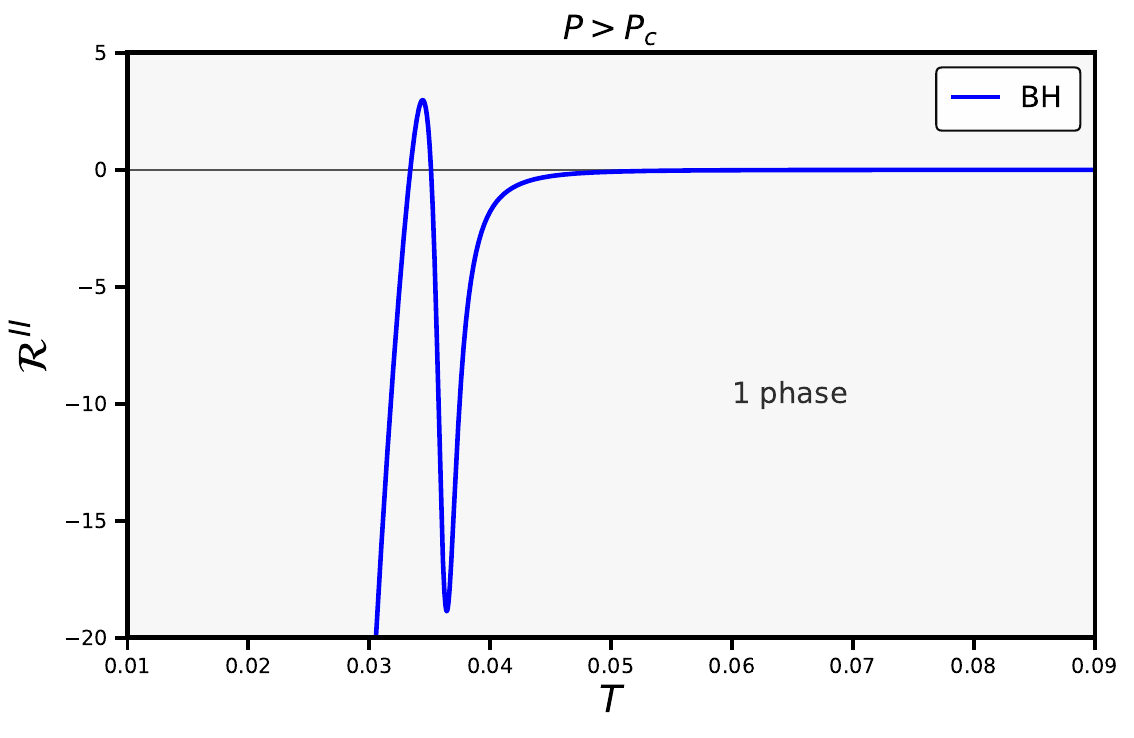}
\end{minipage}
\caption{
Two-dimensional GTD scalar curvature $\mathcal{R}^{II}$ 
as a function of the temperature $T$ for fixed 
$\alpha=1$, $\beta=0.01$, $\gamma=0.2$, and $\lambda=-0.01$. 
The three panels correspond to the subcritical ($P<P_c$), 
critical ($P=P_c$), and supercritical ($P>P_c$) regimes, 
with $P_c\simeq 0.0021$. 
In the subcritical region, the curvature exhibits two divergences 
associated with the SBH/LBH phase transition; 
at the critical point these merge into a single critical divergence, 
while in the supercritical regime the scalar remains regular.
}
\label{fig:GTD scalars tempe}
\end{figure}
In addition, in the subcritical regime $P<P_c$, an intermediate IBH branch appears within the temperature range $T_{\min}<T<T_{\max}$, where the scalar curvature becomes multivalued (see Fig.~\ref{fig:GTD scalars tempe}(a)). This behavior signals the phase coexistence region between SBH and LBH. At the critical point $P=P_c$, the two singularities merge, the IBH branch disappears, and the multivalued structure collapses into a single critical configuration (Fig.~\ref{fig:GTD scalars tempe}(b)). In the LBH branch, the curvature remains negative ($\mathcal{R}^{II}<0$), indicating predominantly attractive interactions, and tends toward zero along this branch, signaling an ideal-gas–like regime with weak thermodynamic interactions. In contrast, the SBH branch typically exhibits positive curvature ($\mathcal{R}^{II}>0$), reflecting dominant repulsive interactions and a more strongly interacting microscopic structure. Hence, the GTD description not only captures the macroscopic phase structure but also reveals a clear qualitative distinction between the microstructures of the SBH and LBH phases.

\vspace{0.1in}

\section{Conclusions}

In this work, we have investigated the optical and thermodynamic properties of a deformed Schwarzschild-AdS black hole surrounded by a cloud of strings and embedded in a perfect fluid dark matter background. Our analysis combines null geodesic techniques, extended black hole thermodynamics, and geometrothermodynamics, providing a unified description of both observable and thermodynamic features of the system.

From the optical perspective, we analyzed the effective potential governing null geodesics and determined the photon sphere radius and the corresponding black hole shadow. The structure of $V_{\rm eff}$ reveals a clear interplay among the parameters of the theory. The PFDM parameter $\lambda$ strengthens the gravitational potential barrier by increasing the height of the effective potential peak, thereby enhancing photon confinement near the unstable circular orbit. In contrast, the cloud of strings parameter $\gamma$ produces a global reduction of the effective potential due to the associated solid angle deficit, lowering the peak and weakening the barrier. The deformation parameter $\beta$ modifies the potential at intermediate scales, smoothing the profile of $V_{\rm eff}$ and shifting the position of its maximum without introducing qualitatively new features. These modifications directly affect the location of the photon sphere and, consequently, the shadow radius. Our numerical analysis shows that the photon sphere radius increases with both $\gamma$ and the magnitude of $\lambda$, for fixed values of the deformation parameter $\beta$. This behavior indicates that the cloud of strings and PFDM contributions systematically enlarge the photon sphere, enhancing the characteristic optical scale of the black hole. In particular, the presence of the cloud of strings introduces a solid angle deficit, leading to a nontrivial relation between the critical impact parameter and the shadow radius, namely $R_{\text{sh}} = \sqrt{1-\gamma}\, b_c$. Overall, the surrounding matter distributions and geometric deformations leave clear imprints on observable quantities, potentially relevant for strong-field astrophysical observations.

Furthermore, on the thermodynamic side, we constructed the quasi-homogeneous fundamental equation and derived the extended first law and generalized Smarr relation. The system exhibits a rich thermal structure, including local extrema of the temperature and a critical point in the extended phase space. However, our analytical perturbative treatment and numerical analysis reveal an important feature: the appearance of van der Waals-type critical behavior is essentially inherited from the RN-AdS sector. Interpreting $\alpha = Q^2$ as the square of an effective Maxwell charge, we find that the critical point $(P_c, v_c, T_c)$ continuously deforms from the standard RN-AdS solution. The corrections induced by the geometric deformation parameter $\beta$, the PFDM parameter $\lambda$, and the string cloud parameter $\gamma$ shift the location of the critical quantities, but do not generate new independent critical structures. In particular, the universal ratio $P_c v_c / T_c$ remains equal to $3/8$ at leading order, and only receives subleading corrections from $\beta$ and $\lambda$. This strongly suggests that the geometrical deformations and the surrounding dark matter field do not favor the emergence of novel phase transitions. Instead, critical behavior is observed only in the vicinity of the RN-AdS configuration, where $\alpha$ plays the role of an effective charge parameter. In addition, we analyzed the leading correction to the Hawking-Page transition induced by the deformation and matter parameters. We found a competing behavior among the couplings: the CS and PFDM contributions increase the Hawking-Page temperature, whereas the parameter $\alpha$ reduces it. As a consequence, the net shift of $T_{HP}$ is governed by the competition between the enhancing effects of $\gamma$ and $\lambda$ and the suppressing role of $\alpha$. This interplay highlights how the surrounding matter sector and the effective charge parameter influence the global thermodynamic stability of the system.

Finally, the geometrothermodynamic analysis provides a consistent geometric characterization of the phase structure. The curvature scalar of the GTD metric diverges exactly at the points where the heat capacity becomes singular, confirming that curvature singularities encode genuine phase transitions. In the subcritical regime $P<P_c$, the SBH branch exhibits larger curvature magnitudes, indicating stronger thermodynamic interactions among its microscopic degrees of freedom. In contrast, the LBH branch is characterized by curvature values close to zero, consistent with a weakly interacting system approaching ideal-gas behavior. This contrast supports the interpretation of the SBH-LBH transition as a change in the strength of microscopic interactions, naturally captured within the GTD framework.

In summary, while the cloud of strings, PFDM, and deformation parameters significantly modify the optical properties and quantitatively shift thermodynamic quantities, they do not qualitatively alter the phase structure from the RN-AdS solution. The presence of critical phenomena remains tightly linked to the effective charge parameter $\alpha$, indicating that the underlying mechanism driving phase transitions is fundamentally electromagnetic-like rather than induced by the surrounding dark matter or geometric deformations.

\footnotesize

\section*{Acknowledgments}

F.A. acknowledges the Inter University Centre for Astronomy and Astrophysics (IUCAA), Pune, India for granting visiting associateship. CRF acknowledge support from Conahcyt-Mexico, grant No. 4003366. The work of HQ was partially supported by UNAM-DGAPA-PAPIIT, grant No. 108225, and SECIHTI, grant No. CBF-2025-I-243.


\begin{thebibliography}{}

\bibitem{aa6} B. P. Abbott et al. (LIGO Scientific Collaboration and Virgo Collaboration),Phys. Rev. Lett. \textbf{116}, 061102 (2016).

\bibitem{aa7} B. P. Abbott et al. (LIGO Scientific Collaboration and Virgo Collaboration),Phys. Rev. Lett. \textbf{119}, 161101 (2017).

\bibitem{aa8} B. P. Abbott et al. (LIGO Scientific Collaboration and Virgo Collaboration),Phys. Rev. Lett. \textbf{116}, 241103 (2016).

\bibitem{aa9} Event Horizon Telescope Collaboration et al.,Astrophys. J. Lett. \textbf{875}, L1 (2019).

\bibitem{aa10} Event Horizon Telescope Collaboration et al.,Astrophys. J. Lett. \textbf{875}, L4 (2019).

\bibitem{aa11} Event Horizon Telescope Collaboration et al.,Astrophys. J. Lett. \textbf{875}, L6 (2019).

\bibitem{aa12} Event Horizon Telescope Collaboration et al.,Astrophys. J. Lett. \textbf{930}, L12 (2022).

\bibitem{aa13} Event Horizon Telescope Collaboration et al., Astrophys. J. Lett. \textbf{930}, L16 (2022).

\bibitem{aa14} Event Horizon Telescope Collaboration et al., Astrophys. J. Lett. \textbf{930}, L17 (2022).

\bibitem{aa1} J. D. Bekenstein, Lett. Nuovo Cim. \textbf{4}, 737 (1972).

\bibitem{aa2} J. D. Bekenstein, Phys. Rev. D \textbf{7}, 2333 (1973). 

\bibitem{aa3} S. W. Hawking, Commun. Math. Phys. {\bf 43}, 199 (1975)

\bibitem{aa4} J. M. Bardeen, B. Carter, and S. W. Hawking, Commun. Math. Phys. \textbf{31}, 161 (1973).

\bibitem{aa15} D. Kastor, S. Ray, and J. Traschen, Class. Quantum Grav. \textbf{26}, 195011 (2009).

\bibitem{aa16} D. Kubiz\v{n}\'ak and R. B. Mann, JHEP \textbf{07}, 033 (2012).

\bibitem{aa17} R.-G. Cai, L.-M. Cao, L. Li, and R.-Q. Yang, JHEP \textbf{09}, 005 (2013).

\bibitem{aa18} A. Sahay, T. Sarkar, and G. Sengupta, JHEP \textbf{07}, 082 (2010).

\bibitem{aa19} S. W. Hawking and D. N. Page, Commun. Math. Phys. \textbf{87}, 577 (1983).

\bibitem{aa20} A. Chamblin, R. Emparan, C. V. Johnson, and R. C. Myers, Phys. Rev. D \textbf{60}, 064018 (1999).

\bibitem{PSL1979} P. S. Letelier, Phys. Rev. {\bf D 20}, 1294 (1979).

\bibitem{aa21} J. P. Morais Graca, I. P. Lobo, V. B. Bezerra and H. Moradpour, Eur. Phys. J. C {\bf 78}, 823 (2018) 

\bibitem{aa22} T. K. Dey, Int. J. Mod. Phys. A {\bf 33}, 1850193 (2018) 

\bibitem{CS1} A.~Kumar and S.~G.~Ghosh, Ann. Phys. (NY) \textbf{484}, 170291 (2026)

\bibitem{CS2} A.~Kumar, Q.~Wu, T.~Zhu and S.~G.~Ghosh, arXiv:2508.02768 [gr-qc].

\bibitem{CS3} A.~Sood, A.~Kumar, J.~K.~Singh and S.~G.~Ghosh, Eur. Phys. J. C \textbf{82}, 227 (2022)

\bibitem{CS4} F.~Ahmed, A.~Al-Badawi \textit{et al.}, arXiv:2510.25764 [astro-ph.HE].

\bibitem{CS5} F.~Ahmed, I.~Sakalli, A.~Al-Badawi, arXiv:2511.11792 [gr-qc].

\bibitem{CS6} {\'I}.~Sakall{\'i}, E.~Sucu, A.~Al-Badawi, F.~Ahmed, arXiv:2512.18881 [gr-qc].

\bibitem{CS7} F.~Ahmed, A.~Al-Badawi, İ.~Sakallı, arXiv:2601.10303 [gr-qc].


\bibitem{CS9} F.~Ahmed, E.~O.~Silva, arXiv:2510.16260 [gr-qc].

\bibitem{CS10} F.~Ahmed, E.~O.~Silva, arXiv:2511.21374 [hep-th].

\bibitem{CS11} F.~Ahmed, A.~Bouzenada, and E.~O.~Silva, Eur.\ Phys.\ J.\ C \textbf{85}, 1385 (2025).

\bibitem{aa5} Planck Collaboration (P. A. R. Ade et al.), Astron. Astrophys. \textbf{571}, A1 (2014).

\bibitem{pp1} G.~Bertone and D.~Hooper, Rev.\ Mod.\ Phys.\ \textbf{90}, 045002 (2018).

\bibitem{pp2} D.~Huterer and D.~L.~Shafer, Rep.\ Prog.\ Phys.\ \textbf{81}, 016901 (2018).

\bibitem{pp3} F.~Rahaman, K.~K.~Nandi, A.~Bhadra, M.~Kalam, and K.~Chakraborty, Phys.\ Lett.\ B \textbf{694}, 10 (2010).

\bibitem{pp4} A.~A.~Potapov, G.~M.~Garipova, and K.~K.~Nandi, Phys.\ Lett.\ B \textbf{753}, 140 (2016).

\bibitem{pp5} V.~C.~Rubin and W.~K.~Ford, Astrophys.\ J.\ \textbf{159}, 379 (1970).

\bibitem{pp6} M.~S.~Roberts and A.~H.~Rots, Astron.\ Astrophys.\ \textbf{26}, 483 (1973).

\bibitem{pp7} V.~C.~Rubin, W.~K.~Ford Jr., and N.~Thonnard, Astrophys.\ J.\ \textbf{238}, 471 (1980).
 
\bibitem{pp8} D.~Clowe, M.~Bradac, A.~H.~Gonzalez, M.~Markevitch, S.~W.~Randall, C.~Jones, and D.~Zaritsky, Astrophys.\ J.\ \textbf{648}, L109 (2006).

\bibitem{pp9} R.~Mandelbaum \textit{et al.}, Mon.\ Not.\ R.\ Astron.\ Soc.\ \textbf{432}, 1544 (2013).

\bibitem{pp10} G.~Hinshaw \textit{et al.} (WMAP Collaboration), Astrophys.\ J.\ Suppl.\ \textbf{208}, 19 (2013).

\bibitem{pp11} P.~A.~R.~Ade \textit{et al.} (Planck Collaboration), Astron.\ Astrophys.\ \textbf{571}, A16 (2014).

\bibitem{pp12} M.~Davis, G.~Efstathiou, C.~S.~Frenk, and S.~D.~M.~White, Astrophys.\ J.\ \textbf{292}, 371 (1985).

\bibitem{VVK} V. V. Kiselev, arXiv: gr-qc / 0303031. 

\bibitem{MHL2012} M. H. Li, and K. C. Yang, Phys. Rev. {\bf D 86}, 123015 (2012).

\bibitem{kk1} S.~Haroon, M.~Jamil, K.~Jusufi, and R.~B.~Mann, Phys.\ Rev.\ D \textbf{99}, 044015 (2019).

\bibitem{kk2} M.~Rizwan, M.~Jamil, and K.~Jusufi, Phys.\ Rev.\ D \textbf{99}, 024050 (2019).

\bibitem{kk3} S.~H.~Hendi, A.~Nemati, K.~Lin, and M.~Jamil, Eur.\ Phys.\ J.\ C \textbf{80}, 296 (2020).

\bibitem{kk4} F.~Atamurotov, U.~Papnoi, and K.~Jusufi, Class.\ Quantum Grav.\ \textbf{39}, 025014 (2022).

\bibitem{kk6} F.~Atamurotov, A.~Abdujabbarov, and W.-B.~Han, Phys.\ Rev.\ D \textbf{104}, 084015 (2021).

\bibitem{kk7} J.~Rajimbaev, S.~Shaymatov, and M.~Jamil, Eur.\ Phys.\ J.\ C \textbf{81}, 699 (2021).

\bibitem{kk8} G.~Rakhimova, F.~Atamurotov, F.~Javed, A.~Abdujabbarov, and G.~Mustafa, Nucl.\ Phys.\ B \textbf{996}, 116363 (2023).

\bibitem{kk9} F.~Atamurotov, F.~Sarikulov, S.~G.~Ghosh, and G.~Mustafa, Phys.\ Dark Univ.\ \textbf{46}, 101625 (2024).

\bibitem{kk10} B.~Shodikulov, M.~Mirov, F.~Atamurotov, S.~G.~Ghosh, and A.~Abdujabbarov, Phys.\ Dark Univ.\ \textbf{50}, 102096 (2025).

\bibitem{kk11} A.~Sood, M.~S.~Ali, J.~K.~Singh, and S.~G.~Ghosh, Chin.\ Phys.\ C \textbf{48}, 065109 (2024).

\bibitem{kk12} A.~Kumar, A.~Sood, J.~K.~Singh, A.~Beesham, and S.~G.~Ghosh, Phys.\ Dark Univ.\ \textbf{40}, 101220 (2023).

\bibitem{kk13} F.~Ahmed, A.~Al-Badawi, and İ.~Sakallı, arXiv:2509.12264 [gr-qc].

\bibitem{kk14} F.~Ahmed, A.~Al-Badawi, and İ.~Sakallı, arXiv:2602.02621 [gr-qc].

\bibitem{kk15} F.~Ahmed, A.~Al-Badawi, and İ.~Sakallı, arXiv:2602.02586 [gr-qc].

\bibitem{kk16} F.~Ahmed, A.~Al-Badawi, and E.~O.~Silva, arXiv:2602.07806 [gr-qc].

\bibitem{kk17} F.~Ahmed and E.~O.~Silva, arXiv:2510.21948 [gr-qc].

\bibitem{MRK2023} M. R. Khosravipoor and M. Farhoudi,  Eur. Phys. J. C {\bf 83}, 1045 (2023).

\bibitem{DJG2025} D. J. Gogoi, P. Hazarika, J. Bora, and R. Changmai, Fortsch. Phys. {\bf 73}, e70004 (2025).

\bibitem{FA2025} F. Ahmed, A. Al-Badawi, I. Sakalli, and D. J. Gogoi, 	arXiv:2505.12122 [gr-qc].

\bibitem{JYG2025} J.-Y.~Gui, K.-J.~He, H.~Ye, and X.-X.~Zeng, Front.\ Phys.\ \textbf{20}, 025202 (2025).

\bibitem{DL2025} D.~Li, S.~Hu, C.~Deng, and X.~Wu, Fortschr.\ Phys.\ \textbf{73}, e70062 (2025), \url{https://doi.org/10.1002/prop.70062}.

\bibitem{PDU2025} F.~Ahmed, A.~Al-Badawi, and I.~Sakalli, Phys.\ Dark Univ.\ \textbf{48}, 101925 (2025).


\bibitem{JLS1960} J. L. Synge, \textit{Relativity: the general theory} (Interscience Publishers, New York, 1960).

\bibitem{AS2024} A. Sood, Md Sabir Ali, J. K. Singh and S. G. Ghosh, Chin. Phys. C {\bf 48}, 065109 (2024).

\bibitem{Singh2025} D.~V.~Singh, S.~Upadhyay, Y.~Myrzakulov, K.~Myrzakulov, B.~Singh and M.~Kumar, Nucl. Phys. B \textbf{1016}, 116915 (2025). 

\bibitem{SC1984} S. Chandrasekhar, {\it The Mathematical Theory of Black Holes} (Oxford University Press, Oxford, 1984).

\bibitem{RMW1984} R. M. Wald, \textit{General Relativity}, University of Chicago Press, Chicago (1984).

\bibitem{Volker2022} V. Perlick and O. Yu. Tsupko, Phys. Rep. {\bf 947}, 1 (2022).

\bibitem{kubizvnak2017black} D.~Kubiz\v{n}\'ak, R.~B.~Mann and M.~Teo, Class.\ Quant.\ Grav.\ \textbf{34}, no.6, 063001 (2017).

\bibitem{romero2024extended1} C.~E.~Romero-Figueroa and H.~Quevedo, \textit{Eur. Phys. J. C} \textbf{84}, 1091 (2024).

\bibitem{romero2026quasi}
C.~E.~Romero-Figueroa and H.~Quevedo,
arXiv:2601.04639 [gr-qc] (2026).

\bibitem{ladino2025phase} J.~M.~Ladino, C.~E.~Romero-Figueroa and H.~Quevedo, \textit{Nucl. Phys. B} \textbf{1018}, 117031 (2025).

\bibitem{ladino2024phase} J.~M.~Ladino, C.~E.~Romero-Figueroa and H.~Quevedo, \textit{Nucl. Phys. B} \textbf{1009}, 116734 (2024).

\bibitem{gunasekaran2012extended} S.~Gunasekaran, R.~B.~Mann and D.~Kubiz\v{n}\'ak, JHEP {\bf 11}, 110 (2012).


\bibitem{herrera2023anisotropic} A.~Herrera-Aguilar, J.~A.~Herrera-Mendoza, D.~F.~Higuita-Borja, J.~A.~M\'endez-Zavaleta and C.~E.~Romero-Figueroa, Eur.\ Phys.\ J.\ C {\bf 83}, 334 (2023).

\bibitem{herrera2021hyperscaling} 
A.~Herrera-Aguilar, J.~E.~Paschalis and C.~E.~Romero-Figueroa, 
arXiv:2110.04445 [hep-th] (2021).

\bibitem{quevedo2007geometrothermodynamics}
H.~Quevedo, J.\ Math.\ Phys.\ {\bf 48}, 013506 (2007).

\bibitem{quevedo2023unified}
H.~Quevedo and M.~N.~Quevedo, Phys.\ Lett.\ B {\bf 838}, 137678 (2023).

\end{thebibliography}
\end{document}